\documentclass[aps,reprint,twocolumn,floatfix,showpacs,pra,superscriptaddress]{revtex4-1}
\usepackage[colorlinks,linkcolor=blue,anchorcolor=blue,citecolor=blue,filecolor=blue,menucolor=blue,runcolor=blue,urlcolor=blue,frenchlinks=blue]{hyperref}
\usepackage{graphicx}
\usepackage{subfigure}
\usepackage{amsmath}
\usepackage{amssymb}
\usepackage{dcolumn}
\usepackage{mathrsfs}
\usepackage{bm,physics}
\usepackage{color,amsfonts,amsthm,mathtools}
\usepackage{subfigure}
\usepackage{setspace}
\usepackage{bm}

\makeatletter

\newcommand{\Rmnum}[1]{\expandafter\@slowromancap\romannumeral #1@}
\makeatother

\begin{document}
\title{Synthetic spin-orbit coupling for the multispin models in optical lattices}

\author{Zhen Zheng}\email{zhenzhen@m.scnu.edu.cn}
\affiliation{Key Laboratory of Atomic and Subatomic Structure and Quantum Control (Ministry of Education), Guangdong Basic Research Center of Excellence for Structure and Fundamental Interactions of Matter, School of Physics, South China Normal University, Guangzhou 510006, China}
\affiliation{Guangdong Provincial Key Laboratory of Quantum Engineering and Quantum Materials, Guangdong-Hong Kong Joint Laboratory of Quantum Matter, Frontier Research Institute for Physics, South China Normal University, Guangzhou 510006, China}
\affiliation{Department of Physics and HK Institute of Quantum Science \& Technology, The University of Hong Kong, Pokfulam Road, Hong Kong, China}

\author{Yan-Qing Zhu}
\affiliation{Department of Physics and HK Institute of Quantum Science \& Technology, The University of Hong Kong, Pokfulam Road, Hong Kong, China}
\affiliation{Quantum Science Center of Guangdong-Hong Kong-Macao Greater Bay Area, Shenzhen, China}

\author{Shanchao Zhang}
\affiliation{Key Laboratory of Atomic and Subatomic Structure and Quantum Control (Ministry of Education), Guangdong Basic Research Center of Excellence for Structure and Fundamental Interactions of Matter, School of Physics, South China Normal University, Guangzhou 510006, China}
\affiliation{Guangdong Provincial Key Laboratory of Quantum Engineering and Quantum Materials, Guangdong-Hong Kong Joint Laboratory of Quantum Matter, Frontier Research Institute for Physics, South China Normal University, Guangzhou 510006, China}
%\affiliation{Quantum Science Center of Guangdong-Hong Kong-Macao Greater Bay Area, Shenzhen, China}

\author{Shi-Liang Zhu}\email{slzhu@scnu.edu.cn}
\affiliation{Key Laboratory of Atomic and Subatomic Structure and Quantum Control (Ministry of Education), Guangdong Basic Research Center of Excellence for Structure and Fundamental Interactions of Matter, School of Physics, South China Normal University, Guangzhou 510006, China}
\affiliation{Guangdong Provincial Key Laboratory of Quantum Engineering and Quantum Materials, Guangdong-Hong Kong Joint Laboratory of Quantum Matter, Frontier Research Institute for Physics, South China Normal University, Guangzhou 510006, China}
\affiliation{Quantum Science Center of Guangdong-Hong Kong-Macao Greater Bay Area, Shenzhen, China}

\author{Z. D. Wang}\email{zwang@hku.hk}
\affiliation{Department of Physics and HK Institute of Quantum Science \& Technology, The University of Hong Kong, Pokfulam Road, Hong Kong, China}
\affiliation{Quantum Science Center of Guangdong-Hong Kong-Macao Greater Bay Area, Shenzhen, China}

%----------------------------------------------------------------------------------------
\begin{abstract}

The essential role of synthetic spin-orbit coupling in discovering new topological matter phases with cold atoms is widely acknowledged. However, the engineering of spin-orbit coupling remains unclear for arbitrary-spin models due to the complexity of spin matrices. In this paper, we develop a more general but relatively straightforward method to achieve spin-orbit coupling for multispin models. Our approach hinges on controlling the coupling between distinct pseudo-spins through two intermediary states, resulting in tunneling with spin flips that have direction-dependent strength. The engineered spin-orbit coupling can facilitate topological phase transitions with Chern numbers over 1, a unique characteristic of multispin models compared to spin-1/2 models. By utilizing existing cold atom techniques, our proposed method provides an ideal platform for investigating topological properties related to large Chern numbers.

\end{abstract}
\maketitle
%----------------------------------------------------------------------------------------

\section{Introduction}

Cold atoms exhibit quantum properties and statistics similar to traditional condensed-matter systems,
yet confer the technical advantages of the artificial controllability \cite{Zhang2018Oct-topo-review,Dalibard2011Nov-artificial-field-review,Bloch2012Apr-simu-review,Gross2017Sep-simu-review}.
It has been established as a clean and tunable platform to quantum simulate a variety of physics models that have attracted tremendous interests in cold atomic research of recent decades.
Notably, the successful engineering and realization of the synthetic gauge fields using cold atoms \cite{Goldman2014Nov-artificial-field-review,Eckardt2017Mar-artificial-field-review} has shed light on exploring topological matters~\cite{Hasan2010Nov-topo-review,Qi2011Oct-topo-review} with cold atoms\cite{Dalibard2011Nov-artificial-field-review,Zhang2018Oct-topo-review}, such as atomic spin Hall effect \cite{Beeler2013,Zhu2006Dec-soc-appli,XJLiu2007,Kennedy2013,Lv2021Sep-soc-exp-Rb87}, quantum anomalous Hall effect~\cite{Jotzu2014,LBShao2008,CJWu2008,SLiu2018,Vasic2015,DLDeng2014,Anisimovas2014}, Dirac/Weyl/Majorana fermions~\cite{SLZhu2007,Tarruell2012,Gomes2012,Duca2015,Zhang2015Jul-soc-appli,SLZhu2011,CLQu2015,YWGuo2019}, and topological vacua of Yang-Mills fields~\cite{JZLi2022}.
Among these previous studies, the synthetic spin-orbit coupling (SOC) between atoms plays a crucial role because it provides a direct and simple routine to introduce nontrivial Berry curvature to the system \cite{Galitski2013-soc-review,Zhang2013Dec-soc-review,Zhai2015Feb-soc-review,Zhang2019May-soc-review},
which becomes the origin of the emergent topological phases~\cite{,Dalibard2011Nov-artificial-field-review,Zhang2018Oct-topo-review}.

In neutral atomic gases, SOC is proposed via a two-photon Raman process \cite{Osterloh2005Jun-soc-theo,Ruseckas2005Jun-soc-theo,Zhu2006Dec-soc-appli}.
The atoms can acquire a net momentum transfer when transiting between the coupled internal states.
If one interprets these internal states as the pseudo-spins,
this transition will exhibit the form of an intrinsic gauge field that can emulate the SOC studied in condensed-matter physics.
Experimental investigations have reported the successful synthesization of SOC in atomic species including $^{87}$Rb \cite{Lin2011Mar-soc-exp-Rb87,Hamner2014Jun-soc-exp-Rb87,Olson2014Jul-soc-exp-Rb87,Luo2016Jan-soc-exp-Rb87,Wu2016Oct-soc-exp-Rb87,Li2016Oct-soc-exp-Rb87,An2017Apr-soc-exp-Rb87,Li2017Mar-soc-exp-Rb87,Sun2018Oct-soc-exp-Rb87,Wang2021Apr-soc-exp-Rb87,Lv2021Sep-soc-exp-Rb87}, $^{40}$K \cite{Huang2016Jun-nat-phys-soc-exp-K40,Meng2016Dec-soc-exp-K40}, $^{6}$Li \cite{Cheuk2012Aug-soc-exp-Li6}, $^{87}$Sr \cite{Kolkowitz2017Feb-soc-exp-Sr87,Aeppli2022Oct-soc-exp-Sr87,Liang2023Jan-soc-exp-Sr87}, $^{161}$Dy \cite{Burdick2016Aug-soc-Dy161}, $^{173}$Yb \cite{Livi2016Nov-soc-exp-Yb173,Song2016Dec-soc-exp-Yb173,Song2019Sep-soc-exp-Yb173} but not limited to these.
Most of those achievements \cite{Liu2014Feb-soc-theo,Struck2014Sep-soc-theo,Wall2016Jan-soc-theo} and relative applications \cite{Zhu2006Dec-soc-appli,Zhang2008Oct-soc-appli,Liu2009Jan-soc-appli,Wang2010Oct-soc-appli,Gong2011Nov-soc-appli,DWZhang2012-soc-appli,Ozawa2013Feb-soc-appli,Zheng2013Mar-soc-appli,Deng2014Apr-soc-appli,Xu2015Mar-soc-appli,Zhang2015Jul-soc-appli,Sun2015Oct-soc-appli,Deng2017Feb-soc-appli,Zhou2017Aug-soc-appli,Chen2018Jul-soc-appli,Tang2018Sep-soc-appli,Li2019Jan-soc-appli,Lin2019Apr-soc-appli,Zheng2019Nov-soc-appli,Chen2020Jan-soc-appli,Shen2020Jan-soc-appli,Zhu2020Apr-soc-appli,Sanchez-Baena2020Apr-soc-appli,Chen2022Aug-soc-appli,Zhang2022Oct-soc-appli,Zhu2022Nov-soc-appli,Li2023Jan-soc-appli,Xiong2023Dec-soc-appli,Wang2024Aug-soc-appli} focus on the spin-1/2 model, which is composed of two pseudo-spins.
Hence the reported topological phases are usually characterized by Chern number that does not exceed 1.
Beyond the spin-1/2 model, the multispin ones can bring in more topological properties.
The protocol for synthesizing SOC is thus desirable for exploring these systems, but only the spin-1 model is investigated to date \cite{Campbell2016Mar-soc-spin-1,Mai2018Nov-soc-spin-1,Hou2020May-soc-spin-1,Adhikari2021Jan-soc-spin-1,Cabedo2021Sep-soc-spin-1,Chen2022Aug-soc-spin-1,Su2023Feb-soc-spin-1,Banger2023Oct-soc-spin-1,Xu2024Jun-soc-spin-1,Gangwar2024Apr-soc-spin-1}.
An intuitive scheme is to add more optical fields to couple additional pseudo-spins.
However, the complexity of spin matrices in multispin systems frustrates the manipulations to these fields, hence it remains elusive so far in obtaining the desired SOC.

Here in this paper, we propose a general scheme for realizing SOC in systems with arbitrary pseudo-spins trapped in optical lattices.
The proposal is based on involving two inter-mediate states that bridge the transition between different pseudo-spins, leading to an effective gauge field whose strength is locked to the tunneling direction.
As a result, the SOC is generated simply by one optical field without introducing additional fields.
After applying SOC in the two-dimensional (2D) system, we find rich topological transitions associated with large Chern number, which are a unique feature of the multispin models.
Therefore this proposal can offer a feasible way for exploring and detecting topological properties for the multispin models.

The paper is organized as follows.
In Sec.\ref{sec:model-hamiltonian}, we present the model Hamiltonian of focus.
We show the engineering of the effective SOC by introducing two intermediate states to couple two pseudo-spins.
In Sec.\ref{sec:2d-case}, we extend the results of the one-dimensional (1D) model to the 2D case.
The engineered SOC can exhibit various forms under artificial manipulations.
Such a system hosts topological phases as well as topological edge modes,
and the results are discussed in Sec.\ref{sec:band-topology}.
We discuss the potential experimental implementations of this proposal in Sec.\ref{sec:experiments}.
Finally, we summarize the paper in Sec.\ref{sec:summary}.

\section{Model Hamiltonian} \label{sec:model-hamiltonian}

\begin{figure}[t]
	\centering
	\includegraphics[width=0.48\textwidth]{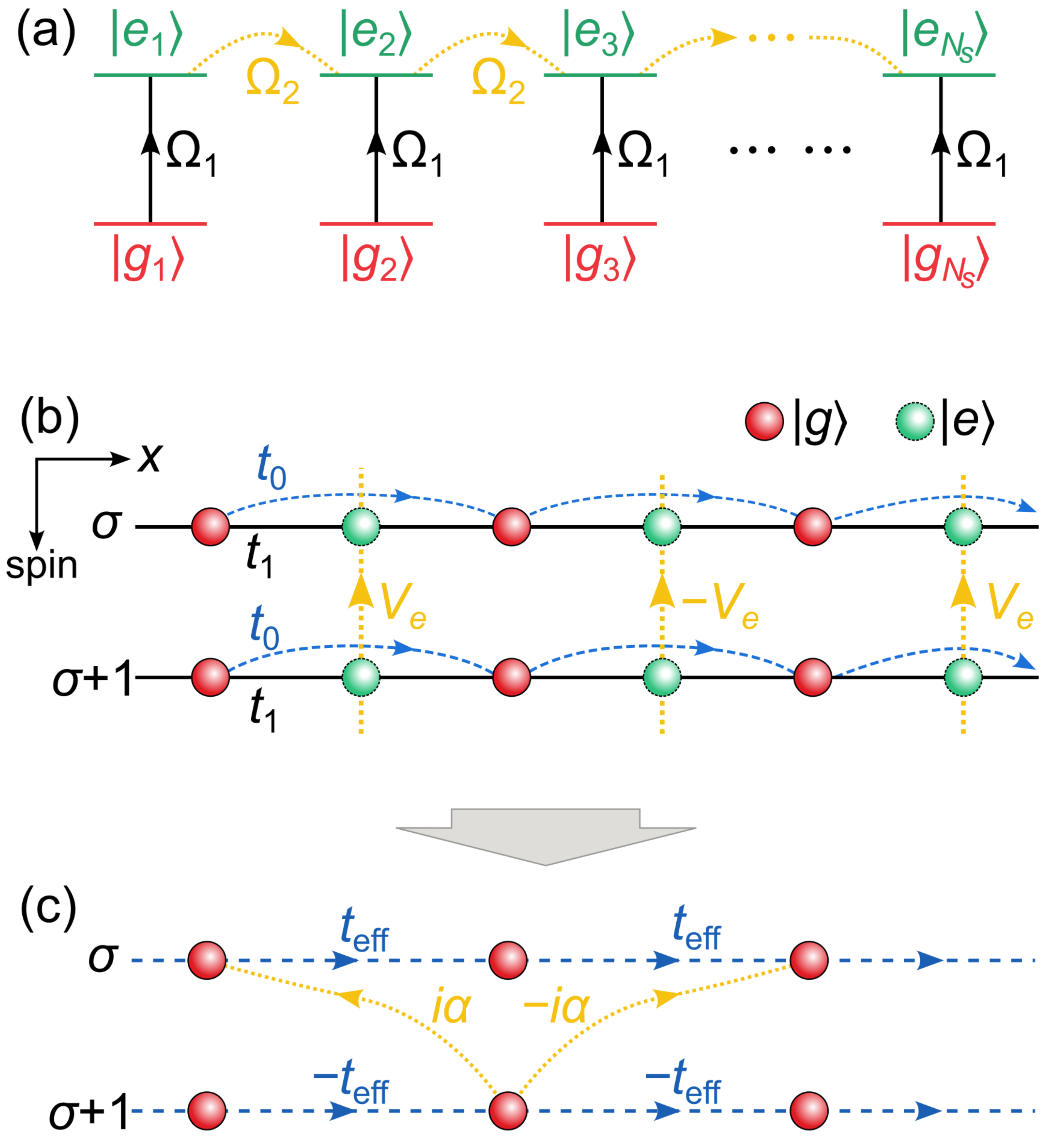}
	\caption{(a) Setups of the atomic level transitions.
	$\ket{g_\sigma}$ and $\ket{e_\sigma}$ are coupled by the field $\Omega_1$ (black solid lines),
	while $\ket{e_\sigma}$ and $\ket{e_{\sigma+1}}$ are coupled by the field $\Omega_2$ (yellow dotted lines).
	(b) Illustration of the 1D lattice model.
	The lattice is composed of $N_s$ chains specified by the pseudo-spin index.
	For each chain, the atoms of $\ket{e_\sigma}$ are loaded in the center of two adjacent atoms of $\ket{g_\sigma}$.
	The transitions driven by $\Omega_1$ in (a) generate intra-chain coupling $t_1$ (black solid lines),
	and the ones driven by $\Omega_2$ generate the inter-chain coupling $V_e$ (yellow dotted lines) but in a staggered pattern in the $x$ direction.
	(c) The mapped lattice model after adiabatically eliminating $\ket{e_\sigma}$.
	It gives rises to effective SOC (yellow dotted lines) for atoms of $\ket{g}$.}
	\label{fig:1d-model}
\end{figure}

We start from the cold atomic gases with two manifolds of internal states.
The atomic level structure is illustrated in Fig.\ref{fig:1d-model}(a).
We label the level manifold with the lower energy as the ground state $\ket{g_{\sigma}}$,
and the other one with the higher energy as the excited state $\ket{e_{\sigma}}$.
Each manifold is composed of pseudo-spins labeled as $\sigma=1,2,\cdots,N_s$ with $N_s$ denoting the number of pseudo-spins.
To establish a simple physics picture of our proposal, we first deal with the 1D case.
The two level manifolds are respectively loaded into two sets of optical lattices with the trap potentials as
$V_{{\rm OL},g}(x) = V_{L}\sin^2(k_Lx)$ and $V_{{\rm OL},e}(x) = V_{L}\cos^2(k_Lx)$.
Here $V_{L}$ denotes the trap depth and $k_L=\pi/d$ with $d$ denoting the lattice constant.
We assume $\hbar=1$ in the whole paper.
Such a setup consists in a superlattice structure shown in Fig.\ref{fig:1d-model}(b),
in which each $\ket{e}$ site resides in the center of two adjacent $\ket{g}$ sites.
Thereby the unit cell is composed of two sites that belong to different sublattices.

We consider the following Hamiltonian,
\begin{equation}
	H=\sum_\sigma\hat{H}_{\sigma}+\hat{H}_{\rm oc} \,. \label{eq:h-start-real-space}
\end{equation}
The first part, i.e. the Hamiltonian of the spin-$\sigma$ atoms, in the free space is expressed as follows,
\begin{align}
	\hat{H}_\sigma &= \int dx\sum_{\lambda=g,e}\psi_{\lambda\sigma}^\dag(x) [ -\frac{\nabla_x^2}{2m} -\mu+ V_{{\rm OL},\lambda}(x) ]\psi_{\lambda\sigma}(x) \notag\\
	&+\Delta_e\psi_{e\sigma}^\dag(x) \psi_{e\sigma}(x) \,. \label{eq:h-0-real-space}
\end{align}
Here $\psi_{\lambda\sigma}$ and $\psi_{\lambda\sigma}^\dag$ stands for the annihilation and creation operators of the $\ket{\lambda_\sigma}$ ($\lambda=g,e$) atoms.
$\mu$ is the chemical potential.
$m$ denotes the atomic mass.
$\Delta_e$ stands for the relative energy detuning of $\ket{e}$ with respect to $\ket{g}$ during the level transition described in the following.

To engineer the Hamiltonian of interest, we generate the optical coupling between the $g$ and $e$ sublattices by using the field $\Omega_1$, which undergoes the transition from $\ket{g_{\sigma}}$ to $\ket{e_{\sigma}}$ within the same pseudo-spin.
A spin-flip term is also introduced via a spatially modulated field $\Omega_2$ that only couples pseudo-spins of $\ket{e}$.
The level transitions are shown in Fig.\ref{fig:1d-model}(a).
It gives the form of the second part of Hamiltonian (\ref{eq:h-start-real-space}),
\begin{equation}
	\hat{H}_{\rm oc} = \int dx\sum_{\sigma=1}^{N_s}\Omega_1 \psi_{e\sigma}^{\dag} \psi_{g\sigma}
	+\sum_{\sigma=1}^{N_s-1} \hat{\Omega}_2(x) \eta_\sigma \psi_{e\sigma}^{\dag} \psi_{e,\sigma+1} +H.c. \label{eq:h-oc-real-space}
\end{equation}
Here $\hat{\Omega}_2(x)$ denotes the spatially modulated mode of the $\Omega_2$ field.
$\eta_\sigma$ is the coefficient dependent on the particular transition from $\ket{e_\sigma}$ to $\ket{e_{\sigma+1}}$ ($\sigma=1,2,\cdots,N_s-1$), and we will discuss its form later.
$H.c.$ stands for the Hermitian conjugation.

To capture the physics of the lattice model, we use the tight-binding approximation,
and expand the atomic operator in terms of the Wannier wave function: $\psi_{\lambda\sigma}(x)=\sum_j W(x-x_{j}^{(\lambda)})\psi_{j\lambda\sigma}$.
Here $x_{j}^{(\lambda)}$ denotes the coordinate of the $j$-th site on the $\lambda$-sublattice,
i.e. $x_{j}^{(g)}=j\times d$ and $x_{j}^{(e)}=x_{j}^{(g)}+d/2$.
Then the model Hamiltonian (\ref{eq:h-start-real-space}) is written as
\begin{equation}
	H = H_0 + H_1 + H_2 \,. \label{eq:h-model-tba}
\end{equation}
The first part of Eq.(\ref{eq:h-model-tba}) originates from Hamiltonian (\ref{eq:h-0-real-space}),
\begin{align}
	H_0 &= -\sum_{\expval{i,j}}\sum_{\lambda,\sigma} t_0 \psi_{i\lambda\sigma}^\dag\psi_{j\lambda\sigma}
	-\sum_{j,\lambda,\sigma} \mu \psi_{j\lambda\sigma}^\dag\psi_{j\lambda\sigma} \notag\\
	&+ \sum_{j,\sigma}\Delta_e\psi_{je\sigma}^\dag\psi_{je\sigma} \,,
	\label{eq:h-0-tba}
\end{align}
where the summation $\sum_{\expval{i,j}}$ is taken over all the nearest-neighbor sites,
and $t_0$ is the tunneling magnitude.

The second part of Eq.(\ref{eq:h-model-tba}) describes the coupling between the $g$ and $e$ sublattices.
Notice that due to the spatial offset between the $g$ and $e$ sublattices,
each $\ket{g}$ atom is coupled to two $\ket{e}$ atoms on its adjacent sites.
Hence $H_1$ is expressed as
\begin{equation}
	H_1 = \sum_{j\sigma}t_1 (\psi_{j,e,\sigma}^\dag +\psi_{j+1,e,\sigma}^\dag )\psi_{j,g,\sigma} + H.c. \,,
	\label{eq:h-1-tba}
\end{equation}
where $t_1=\Omega_1\int  W^*(x+d/2)W(x)dx$.

The last part of Eq.(\ref{eq:h-model-tba}) describes the spin-flip field of the $e$ sublattice.
Practically, its spatial modulated mode can be prepared as $\hat{\Omega}_2(x)=\Omega_2 \cos(k_Lx)$.
Since the Wannier wave function $W(x)$ is of even parity with respect to the site center, $H_2$ is expressed as
\begin{equation}
	H_2 = \sum_{j}\sum_{\sigma=1}^{N_s-1} (-1)^j \hat{V}_e \eta_\sigma \psi_{j,e,\sigma}^\dag \psi_{j,e,\sigma+1} + H.c.\,,
	\label{eq:h-2-tba}
\end{equation}
where the magnitude of the spin-flip field is $\hat{V}_e=\Omega_2 \int \cos(k_Lx)|W(x)|^2 dx$.
Notice that in Eq.(\ref{eq:h-2-tba}), the spatial modulation of the spin-flip term is exhibited as a staggered pattern.
This is ascribed to the $2d$ periodicity of $\hat{\Omega}_2(x)$.
It indicates that the atom will undergo a momentum transfer of $k_L$ when flipping its spin.
Such a modulation plays a crucial role in generating SOC which will be seen later.

In the momentum $k_x$ space, due to the spatial modulation of Hamiltonian (\ref{eq:h-2-tba}),
we choose the basis $\Psi_{k_x}=(\Phi_{k_x,g},\Phi_{k_x,e})^T$, where $\Phi_{k_x,\lambda}=(\psi_{k_x,\lambda,1},\cdots,\psi_{k_x+(\sigma-1)k_L,\lambda,\sigma},\cdots,\psi_{k_x+(N_s-1)k_L,\lambda,N_s})$.
Then Hamiltonian (\ref{eq:h-model-tba}) is recast as
\begin{equation}
	\mathcal{H}(k_x) = \begin{pmatrix}
		\hat{H}_g & \hat{V} \\ \hat{V}^\dag & \hat{H}_e
	\end{pmatrix}\,, \label{eq:h-k-matrix}
\end{equation}
where the block-diagonal terms are
\begin{align*}
	&\hat{H}_g(k_x) = \xi_{k_x}\mathcal{M}_z - \mu \,, \\
	&\hat{H}_e(k_x) = \hat{H}_g + \Delta_e + \hat{V}_e\mathcal{M} + \hat{V}_e^*\mathcal{M}^\dag \,,
\end{align*}
and the block-off-diagonal term $\hat{V}$ is
\begin{align*}
	\hat{V}(k_x) &=
	{\rm Diag}\Big[t_1+t_1e^{ik_xd},\cdots,t_1+t_1e^{i[k_x+(\sigma-1)k_L]d},\\
	&\qquad\cdots,t_1+t_1e^{i[k_x+(N_s-1)k_L]d}\Big] \\
	&= t_1 \times {\rm Diag}\Big[1+e^{ik_xd},\cdots,1+(-1)^{\sigma-1}e^{ik_xd},\\
	&\qquad\cdots,1+(-1)^{N_s-1}e^{ik_xd}\Big] \,.
\end{align*}
Here $\xi_{k_x}=-2t_0\cos(k_xd)$. $\mathcal{M}_z = {\rm Diag}[1,-1,1,-1,\cdots]$.
The nonzero elements of the matrix $\mathcal{M}$ are
\begin{equation}
	\mathcal{M}_{\sigma,\sigma+1}=\eta_\sigma \label{eq:m-elements}
\end{equation}
for $\sigma=1,2,N_s-1$, otherwise $\mathcal{M}_{\sigma,\sigma'}=0$.

We assume that the atoms are initially prepared in $\ket{g}$ and far detuned from $\ket{e}$,
yielding that $\ket{e}$ can be adiabatically eliminated.
The effective Hamiltonian is then obtained by $\mathcal{H}_{\rm eff}(k_x) = \hat{H}_g + \hat{V}(\hat{H}_g-\hat{H}_e)^{-1}\hat{V}^\dag$ from Hamiltonian (\ref{eq:h-k-matrix}) (see Appendix \ref{sec:app:effective-hamiltonian}),
which gives
\begin{equation}
	\mathcal{H}_{\rm eff}(k_x) = \xi_{k_x}\mathcal{M}_z-\mu + \hat{\zeta}_{k_x}
	+ (\hat{\alpha}_{k_x}\hat{V}_e\mathcal{M} + H.c.) \,. \label{eq:h-eff-rare}
\end{equation}
Here $\hat{V}_e$ is set as $\hat{V}_e=V_ee^{i\varphi_e}$ with $\varphi_e$ denoting its phase,
$\hat{\zeta}_{k_x} = {\rm Diag}[\hat{\zeta}_{1}(k_x),\cdots, \hat{\zeta}_{N_s}(k_x)]$, and
\begin{equation*}
	\begin{cases}
		\hat{\zeta}_{\sigma}(k_x)=-\frac{\Delta_e}{\Delta_e^2-V_e^2}t_1^2|1 - (-1)^{\sigma} e^{ik_xd}|^2 \\
		\hat{\alpha}_{k_x}=-\frac{V_ee^{i\varphi_e}}{\Delta_e^2-V_e^2}t_1^2(1+e^{ik_xd})(1-e^{-ik_xd})
	\end{cases} \,.
\end{equation*}
Hamiltonian (\ref{eq:h-eff-rare}) can be further written in a compact form,
\begin{align}
	& \mathcal{H}_{\rm eff}(k_x) = \xi_{\rm eff}(k_x)\mathcal{M}_z-\mu_{\rm eff} + \mathcal{H}_{\rm soc}(k_x) \,, \label{eq:h-1D-case}\\
	& \mathcal{H}_{\rm soc}(k_x) = - 2\alpha V_e \sin(k_xd) (ie^{i\varphi_e}\mathcal{M} + H.c.) \,. \label{eq:soc-form}
\end{align}
Here $\xi_{\rm eff}(k_x)=-2t_{\rm eff}\cos(k_xd)$,
$t_{\rm eff}=t_0+\zeta/2$ with $\zeta=2t_1^2\Delta_e/(\Delta_e^2-V_e^2)$
is the effective tunneling magnitude,
$\mu_{\rm eff}=\mu+\zeta$ is the effective chemical potential,
and
\begin{equation}
	\alpha=\frac{t_1^2 V_e}{\Delta_e^2-V_e^2} \,. \label{eq:soc-strength}
\end{equation}

From Eq.(\ref{eq:h-1D-case}),
we can see that the final effective Hamiltonian is now assigned with the $\alpha$ term.
Such a term describes a nearest-neighbor tunneling process associated with the spin flip,
and the sign of the tunneling magnitude is locked to tunneling direction.
Hence the tunneling process indeed exhibits the physics picture of SOC.
For the spin-1/2 case simply with $\eta_\sigma=1$ in Eq.(\ref{eq:m-elements}), it gives rise to $\mathcal{H}_{\rm soc}=2\alpha \sin(k_xd)\sigma_y$ when $\varphi_e=0$
and $\mathcal{H}_{\rm soc}=-2\alpha \sin(k_xd)\sigma_x$ when $\varphi_e=\pi/2$,
where $\sigma_{x,y,z}$ denotes the Pauli matrix.
For the general case with $N_s$ spins,
the spin matrices will have complex forms beyond the Pauli matrices as $S_z={\rm Diag}[-S,-S+1,\cdots,S]$ with $S=(N_s-1)/2$,
and $S_{x,y}$ are derived by the matrix representation of the raising and lowering operators $S_{\pm}$,
reading $S_x = (S_+ + S_-)/2$ and $S_y = (S_+ - S_-)/(2i)$.
Here the nonzero elements of the $S_{\pm}$ matrices are given by $[S_\pm]_{\sigma \pm 1,\sigma}=\frac{1}{2}\sqrt{S(S+1)-(\sigma-S-1)(\sigma-S)}$.
Therefore, by engineering $\mathcal{M} = S_-$ via the optical method, we can obtain the generalized form of SOC for systems with $N_s$ spins.
Intuitively, one might consider employing multiple optical fields to couple distinct pseudo-spins, each with independently controllable intensities. However, the preparation of an increased number of optical fields will complicate practical implementation and potentially lead to additional heating effects. Consequently, this motivates our exploration of an alternative approach that may simplify the setups of SOC engineering.
We will discuss this issue in Sec.\ref{sec:experiments},
and one will find that when representing the elements of $\mathcal{M}$ by the Clebsch-Gordan coefficients during the transitions between pseudo-spins, the SOC can be engineered without involving additional multiple optical fields.

In Eq.(\ref{eq:soc-form}), the synthetic SOC is composed of two parts as $ie^{i\varphi_e}= i\cos(\varphi_e)-\sin(\varphi_e)$.
It reveals that the two parts have a relative $\pi/4$ phase, and their relative strength of can be controllable by changing $\varphi_e$.
Since it is engineered by adiabatically eliminating two intermediate $\ket{e}$ states,
the engineered SOC indeed originates from a third-order perturbative process with respect to Hamiltonian (\ref{eq:h-0-tba}),
which can be also recognized by the form of Eq.(\ref{eq:soc-strength}).
Therefore the SOC strength depends on not only the $g$-$e$ coupling $t_1$ but also the staggered spin-flip field $V_{e}$.
This facilitates the artificial controllability to the engineered SOC.
We note that in obtaining Eq.(\ref{eq:h-eff-rare}), we have made the assumption of far detuning.
This yields $\Delta_e>V_e$, hence $\alpha$ in Eq.(\ref{eq:soc-strength}) is positive.

\section{2D case}\label{sec:2d-case}

\begin{figure}[t]
	\centering
	\includegraphics[width=0.48\textwidth]{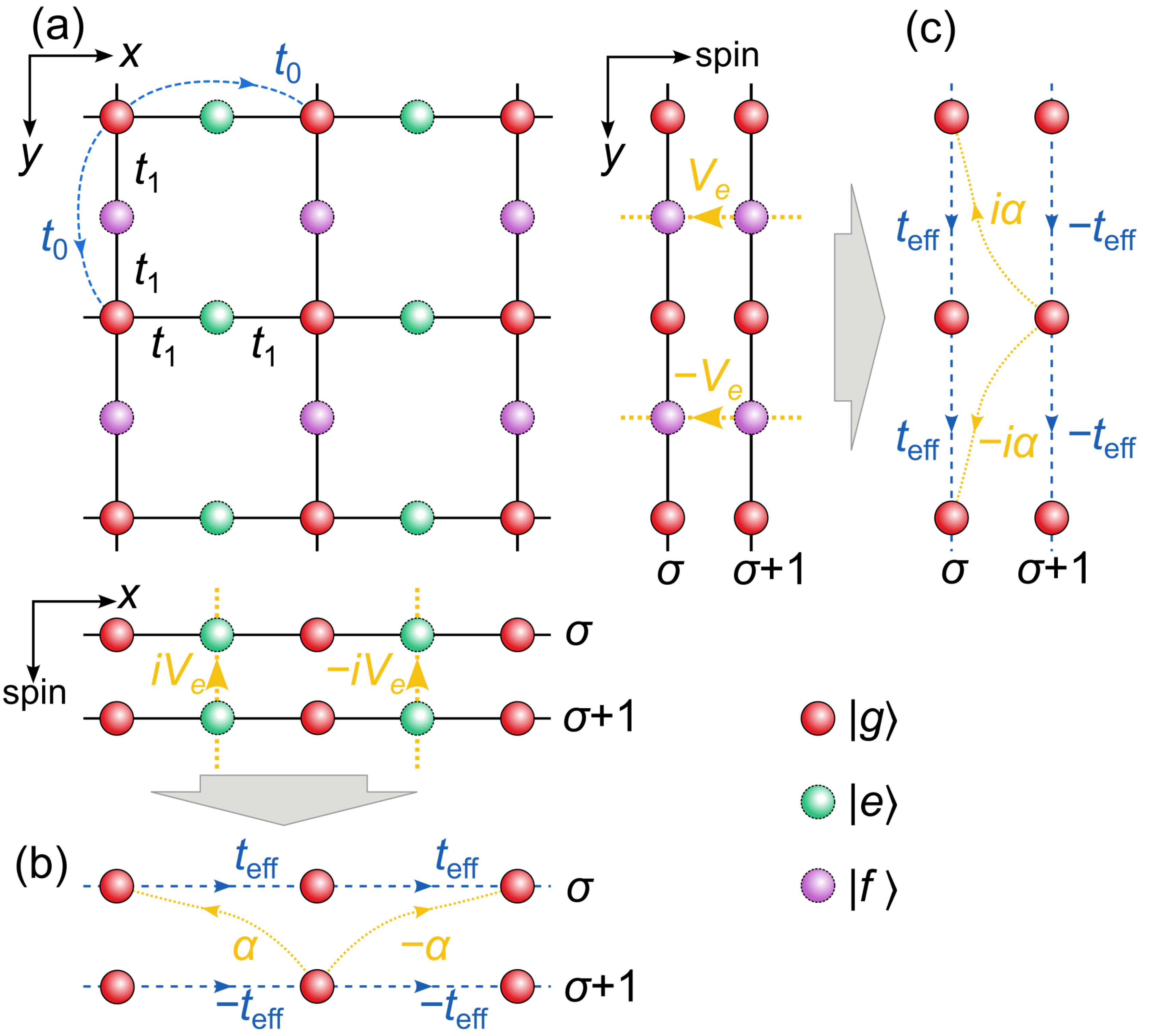}
	\caption{(a) Illustration of the 2D lattice model.
	The lattice is composed of $N_s$ planes specified by the pseudo-spin index.
	For each plane, the atoms of $\ket{e}$ (or $\ket{f}$) are loaded in the center of two adjacent atoms of $\ket{g}$ in the $x$ (or $y$) direction.
	(b)-(c) The mapped lattice models after adiabatically eliminating $\ket{e_\sigma}$ and $\ket{f_\sigma}$, and projecting in the $x$ and $y$ directions.
	It gives rises to effective SOC (yellow dotted lines) of which the strength and phase can be separately controlled in the $x$ and $y$ directions.
	Here we show the example of the Rashba SOC.}
	\label{fig:2d-model}
\end{figure}

The proposal can be extended to the 2D lattice model,
which can result in a variety of SOCs with rich forms.
To individually manipulate SOC in different spatial dimensions,
a third manifold of excited states $\ket{f_\sigma}$ is necessary,
and thereby the extended atomic level transitions are illustrated in Fig.\ref{fig:2d-model}.
The lattice potential is accommodated correspondingly:
the $\ket{e}$ atoms reside in the center of two adjacent $g$ sites along the $x$ direction;
in contrast the $\ket{f}$ atoms reside in a similar pattern but along the $y$ direction.
They lead to a Lieb superlattice structure, which is shown in Fig.\ref{fig:2d-model}(a).

\begin{figure}[t]
	\centering
	\includegraphics[width=0.48\textwidth]{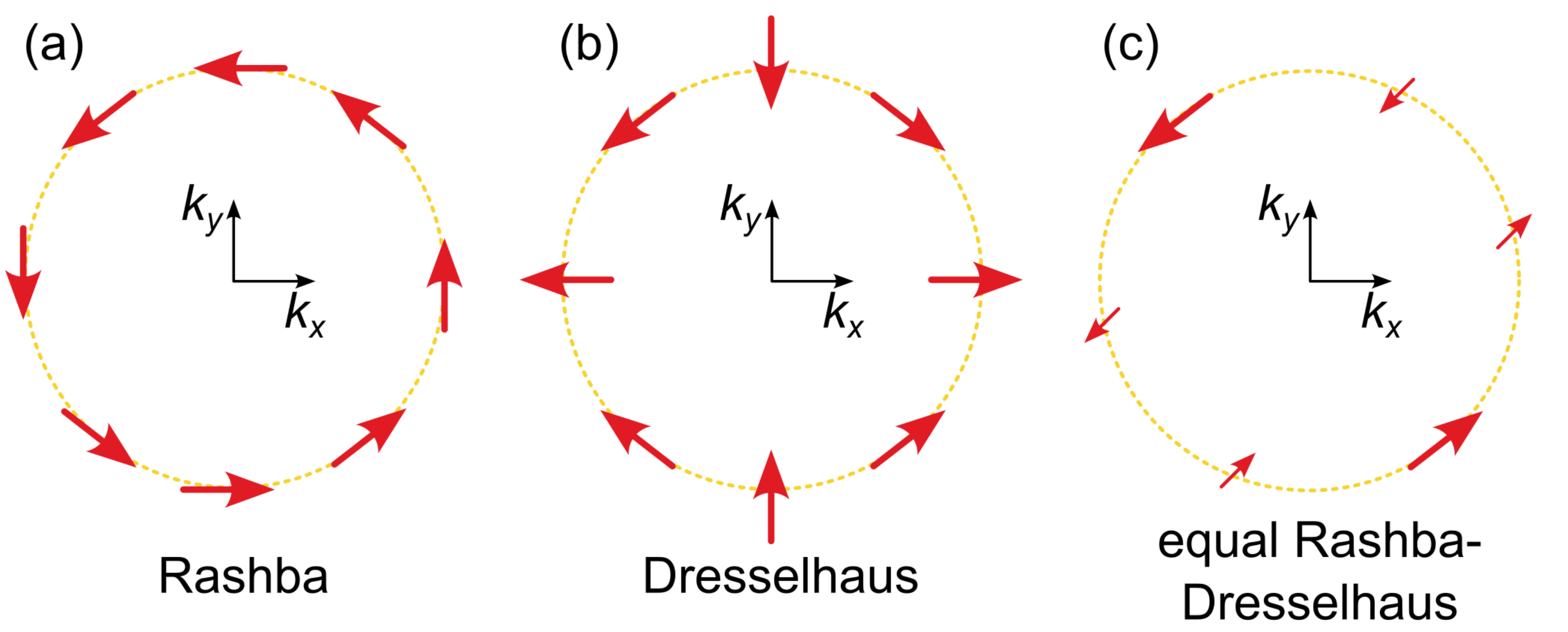}
	\caption{Illustration of the spin texture in the momentum space generated by (a) Rashba, (b) Dresselhaus, and (c) equal Rashba-Dresselhaus SOC. The pseudo-spin orientation is shown by the red arrows, whose length indicates the magnitude.}
	\label{fig:spin-texture}
\end{figure}

We process the same derivations as formulated in Sec.\ref{sec:model-hamiltonian}.
After adiabatically eliminating $\ket{e}$ and $\ket{f}$, the effective Hamiltonian is written in the same form of Eq.(\ref{eq:h-1D-case}) as,
\begin{equation}
	\mathcal{H}_{\rm eff}(\vb{k}) = \xi_{\rm eff}(\vb{k})\mathcal{M}_z-\mu_{\rm eff} + \mathcal{H}_{\rm soc}(\vb{k}) \,, \label{eq:h-2D-case}
\end{equation}
where $\xi_{\rm eff}(\vb{k}) = -2t_{\rm eff}[\cos(k_xd)+\cos(k_yd)]$,
and SOC for the 2D case is presented as
\begin{align}
	\mathcal{H}_{\rm soc}(\vb{k}) &= -i2\alpha V_e \sin(k_xd) e^{i\varphi_e^{(x)}}\mathcal{M} \notag\\
	&-i2\alpha V_e \sin(k_yd) e^{i\varphi_e^{(y)}}\mathcal{M} + H.c. \label{eq:h-soc-2D-case}
\end{align}
Practically, $\varphi_e^{(x)}$ and $\varphi_e^{(y)}$ can be separately tuned.
By preparing $\mathcal{M}=S_-$ as discussed in Sec.\ref{sec:model-hamiltonian},
one can obtain (i) the Rashba SOC \cite{Bychkov1984Nov-rashba-soc}
by setting $(\varphi_e^{(x)},\varphi_e^{(y)})=(0,\pi/2)$,
\begin{equation}
	\mathcal{H}_{\rm soc}^{(R)}(\vb{k}) = 2\alpha V_e [\sin(k_xd) S_y - \sin(k_yd) S_x] \,, \label{eq:h-soc-rashba}
\end{equation}
(ii) the Dresselhaus SOC \cite{Dresselhaus1955Oct-dresselhaus-soc}
by setting $(\varphi_e^{(x)},\varphi_e^{(y)})=(-\pi/2,\pi)$,
\begin{equation}
	\mathcal{H}_{\rm soc}^{(D)}(\vb{k}) = 2\alpha V_e [\sin(k_xd) S_x - \sin(k_yd) S_y] \,, \label{eq:h-soc-dresselhaus}
\end{equation}
and (iii) the equal Rashba-Dresselhaus SOC by setting $(\varphi_e^{(x)},\varphi_e^{(y)})=(-\pi/4,3\pi/4)$,
\begin{equation}
	\mathcal{H}_{\rm soc}^{(RD)}(\vb{k}) = \frac{1}{\sqrt{2}}[ \mathcal{H}_{\rm soc}^{(R)}(\vb{k}) + \mathcal{H}_{\rm soc}^{(D)}(\vb{k}) ] \,. \label{eq:h-soc-equal-rashba-dresselhaus}
\end{equation}
The above three types of SOC exhibit various patterns in the spin texture, which are illustrated in Fig.\ref{fig:spin-texture}.
Intriguing topological properties can arise from these non-trivial spin texture structures.

\section{Band topology}\label{sec:band-topology}

It has been known that the band inversion accompanying SOC can support the topological phase transition from the trivial to non-trivial gapped phases.
Specifically in the 2D case, this is known to be captured by the quantized Chern number as the topological invariant \cite{Qi2008Nov-topo-ins},
\begin{equation}
	\mathcal{C} = \sum_{n \in \text{occ}} \mathcal{C}_n \,,\label{eq:chern-total-def}
\end{equation}
where the $n$'s summation is taken over all the occupied (abbreviated as occ) bands,
and $\mathcal{C}_n$ is the Chern number of the $n$-th band calculated as
\begin{equation*}
	\mathcal{C}_n = \frac{1}{2\pi}\int_{\rm BZ} \mathcal{A}_n(\vb{k}) d\vb{k} \,.
\end{equation*}
Here BZ stands for the Brillouin zone.
$\mathcal{A}_n(\vb{k})$ is the Berry curvature of the $n$-th band formulated as
\begin{equation*}
	\mathcal{A}_n(\vb{k}) =
	i\sum_{m\neq n}\frac{\bra{n}\nabla_{k_x}\hat{H}\ket{m}\bra{m}\nabla_{k_y}\hat{H}\ket{n}}{E_n(\vb{k})-E_m(\vb{k})} - (k_x\leftrightarrow k_y) \,.
\end{equation*}
$\ket{n}$ and $E_n(\vb{k})$ denote the eigen-state and eigen-energy of the $n$-th band for the Hamiltonian $\hat{H}$.

\begin{figure}[t]
	\centering
	\includegraphics[width=0.48\textwidth]{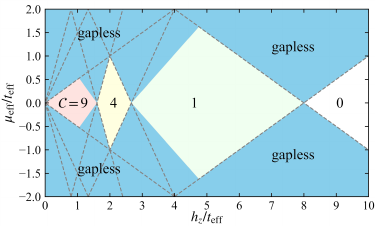}
	\caption{Phase diagram for the 2D model with $N_s=6$ and Rashba SOC.
	The number denotes the Chern number $\mathcal{C}$ of each phase.
	The blue regions outline the gapless phase.
	The gray-dashed lines correspond to the gap-closing condition indicated by Eq.(\ref{eq:gap-closing}).
	We set $\alpha=1.2t_{\rm eff}$.}
	\label{fig:phase-diagram}
\end{figure}

Notice that in the effective Hamiltonian (\ref{eq:h-2D-case}), the matrix $\mathcal{M}_z$ is not identical to $S_z$, and hence the system is gapless.
Since $\mathcal{C}$ is only well defined in a gapped system,
we introduce an additional Zeeman term with the strength $h_z$ into the effective Hamiltonian (\ref{eq:h-2D-case}),
\begin{equation}
	\hat{H}(\vb{k}) = \mathcal{H}_{\rm eff}(\vb{k}) + h_zS_z \,. \label{eq:h-add-hz}
\end{equation}
Hereafter for simplicity without loss of generality, we consider the $N_s=6$ model with the Rashba SOC in $\mathcal{H}_{\rm eff}$, see Eq.(\ref{eq:h-soc-rashba}).
In experiments, this Zeeman term can be naturally present by the Zeeman splitting or the energy offset for the pseudo-spin states.
Generally, SOC induces the nontrivial Berry curvature, yet does not close the band gap \cite{Zhang2018Oct-topo-review}.
Hence in Fig.\ref{fig:phase-diagram} at a fixed $\alpha$, we plot the phase diagram in the $h_z$-$\mu_{\rm eff}$ plane.
At $\mu_{\rm eff}=0$, by increasing $h_z$, $\mathcal{C}$ processes the changes from 9, 4, 1, to 0.
It reveals that the transitions occur within three topological phases with nonzero $\mathcal{C}$.
When $\mu_{\rm eff}\neq 0$, the topological phases are separated by the gapless phase.
We remark that the phenomenon associated with the emergent large $\mathcal{C}$ is the feature of the multispin models that is distinct from the spin-1/2 one \cite{Qi2008Nov-topo-ins},
because $\mathcal{C}$ is limited by its maximum $N_s$.

\begin{figure}[t]
	\centering
	\includegraphics[width=0.48\textwidth]{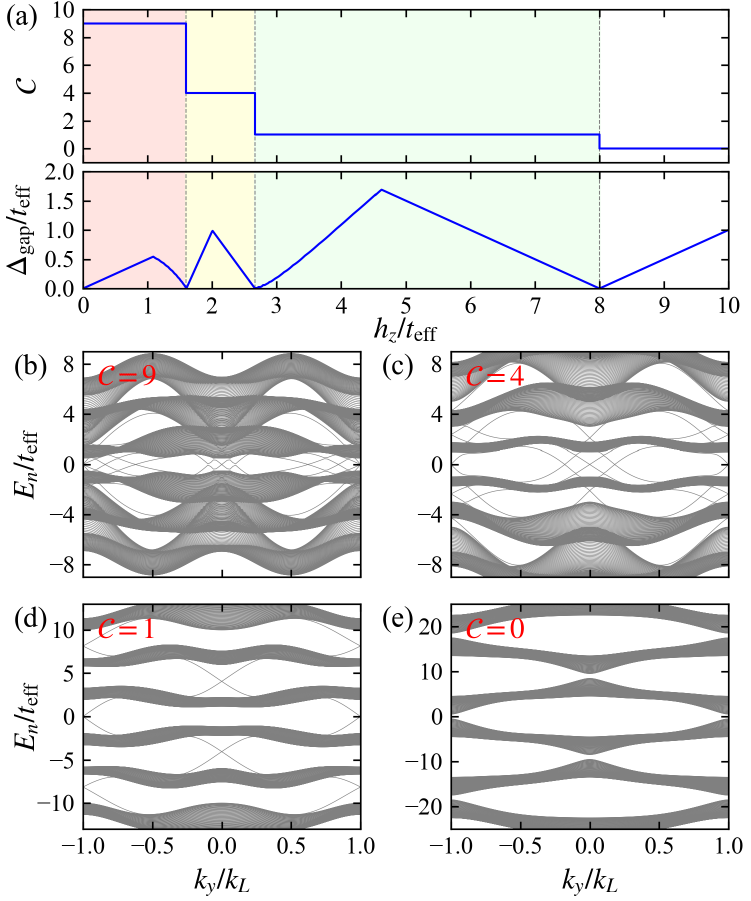}
	\caption{(a) The Chern number $\mathcal{C}$ and gap $\Delta_{\rm gap}$ as functions of $h_z$. The gray dashed lines mark the critical points of the phase transitions. We set $\mu_{\rm eff}=0$ and $\alpha=1.2t_{\rm eff}$.
	(b)-(e) Spectrum of the system with the cylindrical geometry for $h_z=1t_{\rm eff}$ (b), $2t_{\rm eff}$, $4t_{\rm eff}$ (d), and $9t_{\rm eff}$ (e). The value of the Chern number $\mathcal{C}$ is shown in each panel. Other parameters are the same as in (a).}
	\label{fig:topo-trans}
\end{figure}

Generally, the topological phase transitions accompany the closing and reopening of the band gaps.
To extract a clear picture of the topological transitions with large $\mathcal{C}$,
we calculate the gap characterized by the quantity $\Delta_{\rm gap}={\rm min}(|E_n(\vb{k})|)$ and plot it as a function of $h_z$ in Fig.\ref{fig:topo-trans}(a).
We find $\Delta_{\rm gap}$ vanishes at the critical points during the topological phase transitions.
yielding that the band gap closes and reopens.
In Hamiltonian (\ref{eq:h-add-hz}), the band only closes at $\vb{k}_c=(k_c,k_c)$
with $k_c=0,\pm k_L$, i.e. at the center or corners of the BZ.
It indicates that $\alpha$ does not affect the gap closing,
and thus we do not further inspect the influence of its magnitude on the topological transition.
The phase boundaries are intuitively determined by assuming that one of the diagonal terms in Eq.(\ref{eq:h-add-hz}) vanishes. It gives the following equation,
\begin{equation}
	|K_1t_{\rm eff} - \mu_{\rm eff} \pm h_z K_2/2 | = 0 \,, \label{eq:gap-closing}
\end{equation}
where the integers $K_1=0$ or $\pm 4$, and $K_2=1$, 3, or 5.
Particularly at $\mu_{\rm eff}=0$, it gives rise to the phase boundaries at $h_z/t_{\rm eff}=0,\frac{8}{5}, \frac{8}{3},8$.
This result has been verified in Fig.\ref{fig:topo-trans}(a).
When $\mu_{\rm eff}\neq 0$, several phase boundaries deviate from the prediction of Eq.(\ref{eq:gap-closing}), as shown in Fig.\ref{fig:phase-diagram}.
This is because one of six bands is partially filled when increasing $\mu_{\rm eff}$ under the the complex band structure,
hence the system turns to the gapless phase.

As the topological phases support topological edge modes existing in the band gaps,
we plot the spectrum in Fig.\ref{fig:topo-trans}(b)-(e).
We use cylindrical geometry by setting the open boundary condition in the $x$ direction, while periodic boundary condition in the $y$ direction.
According to the bulk-edge correspondence, the number of the existing chiral edge modes is closely tied to $\mathcal{C}$ \cite{Qi2008Nov-topo-ins}.
We calculate $\mathcal{C}_n$ for each band, and obtain $\mathcal{C}_n=\{1,3,5,-5,-3,-1\}$ from the bottom to top bands for panel (b),
$\mathcal{C}_n=\{1,3,0,0,-3,-1\}$ for panel (c),
$\mathcal{C}_n=\{1,0,0,0,0,-1\}$ for panel (d),
and $\mathcal{C}_n=0$ for panel (e).
As defined in Eq.(\ref{eq:chern-total-def}), $\mathcal{C}$ is given by summating $\mathcal{C}_n$ of all the occupied bands.
Therefore, in Fig.\ref{fig:topo-trans}(b)-(d), one can see that the number of zero-energy nodes in the spectrum is identical to $\mathcal{C}$ (noting that the nodes are two-fold degenerate at $k_y= k_L$ in panel (b), and modes at $k_y=\pm k_L$ are equivalent due to the Bloch theorem).

\section{Experimental Implementations}\label{sec:experiments}

\begin{figure}[t]
	\centering
	\includegraphics[width=0.48\textwidth]{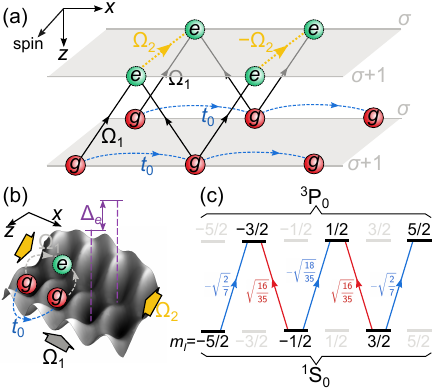}
	\caption{(a) Setups for the experimental realization.
		The atoms are loaded into layers whose index represents the manifold index $\lambda=g,e$ of pseudo-spins.
		Based on this scenario, the field $\Omega_1$ stands for the inter-layer coupling, and $\Omega_2$ stands for the intra-layer coupling.
		(b) Illustration of the model in the lattice trap potential. The detuning $\Delta_e$ is manipulated by the energy offset between sites on adjacent layers. $\Omega_1$ (gray arrows) and $\Omega_2$ (yellow arrows) are respectively placed along the $z$- and $x$-axis.
		(c) Transitions betweens the pseudo-spins for the spin-5/2 model. The blue and red arrows describe the $\sigma^+$ and $\sigma^-$ polarization process. The number beside each arrow indicates the Clebsch-Gordan coefficients during the transitions.}
	\label{fig:exp-setups}
\end{figure}

The proposal for engineering SOC can be realized by existing techniques in cold atoms.
To devise a simplified experimental configuration, we arrange the atoms in layers that are oriented perpendicular to the $z$-axis, as shown in Fig.\ref{fig:exp-setups}(a).
In this way, the manifold index of the pseudo-spins (i.e. $\lambda=g,e$) is represented by the layer index.
Such an index representation has the advantage that it is readily extended to the 2D case,
in which the $\lambda=f$ manifold can be represented by atoms in another layer adjacent to the $g$ layer.
Furthermore, the energy detuning $\Delta_e$ between the two manifolds can be manipulated by the energy offset between the sites on the adjacent layers (e.g. via the Wannier-Stark ladder construct \cite{Gluck2002Aug-wannier-stark-ladder}), as shown in Fig.\ref{fig:exp-setups}(b).
The field $\Omega_1$ is then induced via the inter-layer coupling along the $z$-axis, which can be engineered by the linearly polarized optical field or natural lattice tunneling.

In the above investigations, we use the $N_s=6$ system as a concrete example to present the applications of the SOC in multispin models.
A candidate system that can realize the model is the alkaline-earth-like atomic ensemble of $^{173}$Yb \cite{Livi2016Nov-soc-exp-Yb173},
because its nuclear spin $I=5/2$,
which provides $N_s=2I+1$ nuclear Zeeman states for both the ground $^1$S$_0$ and the meta-stable excited $^3$P$_0$ electronic states.
Likewise for models with higher spins, e.g. the $N_s=10$ one, the candidate atomic species can be chosen as $^{87}$Sr whose nuclear spin $I=9/2$.

As discussed in Sec.\ref{sec:model-hamiltonian}, the key to the synthetic SOC is to prepare $\mathcal{M}$ of Eq.(\ref{eq:m-elements}) to share the form of the lowering operator $S_-$.
In Fig.\ref{fig:exp-setups}(c),
we choose the hyperfine states with $m_I=-5/2,\,-1/2,\,3/2$ of $^1$S$_0$ and $m_I=-3/2,\,1/2,\,5/2$ of $^3$P$_0$ to emulate the pseudo-spins $\sigma=1,3,5,2,4,6$.
Based on this scenario,
the field $\Omega_2$ is produced by $x$-directional lasers with standing-wave modulation and the circular polarization to couple $\ket{e_{\sigma}}$ and $\ket{e_{\sigma \pm 1}}$.
Concurrently, the detuning $\Delta_e$ is utilized to ensure that $\ket{g_{\sigma}}$ is off detuned with respect to $\Omega_2$.
As a result, the coefficients $\eta_\sigma$ in Eq.(\ref{eq:m-elements}) can be represented by the Clebsch-Gordan coefficients during the transitions.
This is sketched in Fig.\ref{fig:exp-setups}(c),
We note that the sign existing in Clebsch-Gordan coefficients can be compensated by preparing a $\pi$ phase between the $\sigma^+$ and $\sigma^-$ polarizations.
Then one can find $|\eta_\sigma| \propto [S_-]_{\sigma-1,\sigma}$, and thus $\mathcal{M}\propto S_-$ up to a multiplicative constant.
We remark that in this way, all the transitions can occur simultaneously by one optical field.
It indicates that, compared with spin-1/2 one, the SOC for the multispin model can be generated without preparing additional fields.

For $^{173}$Yb, we use laser fields of wavelength $\lambda_{\rm laser}=752$nm to construct the optical lattice.
The lattice recoil energy is calculated as $E_R=h^2/(2m\lambda_{\rm laser}^2)\approx 2\pi\hbar\times 8.83$kHz, which we adopt as the energy unit below.
To simplify the estimation of experimental parameters, we apply a 1D lattice model. This model not only facilitates these estimations but also enables us to extrapolate the orders of magnitude for parameters of the 2D case.
When setting the lattice trap depth as $V_L=8E_R$, the natural tunneling magnitude is $t_0\approx 0.0308E_R$ \cite{Walters2013Apr-optical-lattice-cal}.
By adjusting $\Omega_{1,2}$ within the range of 0.5 to 2.0$E_R$ and setting $\Delta_e=1.5E_R$, we obtain the SOC strength $\alpha$ that varies from 0.107 to 1.43$t_{\rm eff}$, from which the parameters in Fig.\ref{fig:phase-diagram} can be achieved.

\section{Conclusion}\label{sec:summary}

In summary, we have presented a scheme of synthesizing SOC for the multispin systems.
A key of this proposal relies on manipulating the coupling of two inter-mediate states,
which leads to a tunneling process between two pseudo-spins
with the strength locked to the tunneling direction.
A variety of SOC can be proposed in this way.
They pave the way for exploring topological phases of the lattice models,
and hence more topological edge modes characterized by the large Chern number may be discovered for such a multispin system.

\begin{acknowledgments}
	This work was supported by National Key Research and Development Program of China (Grant No. 2022YFA1405300), the National Natural Science Foundation of China (Grant No. 12074180), Innovation Program for Quantum Science and Technology (Grant No. 2021ZD0301700), the NSFC/RGC JRS grant (Grant No. N\_HKU 774/21),  GRF (Grants No. 17310622 and No. 17303023) of Hong Kong, and the Science and Technology Program of Guangzhou (Grant No. 2024A04J0004).
\end{acknowledgments}

\appendix

\section{Effective Hamiltonian in the perturbation method}\label{sec:app:effective-hamiltonian}

In this appendix we present the general derivations for obtaining the effective Hamiltonian using the purterbation method \cite{Cohen-Tannoudji1998book}.
We consider a quantum system described by the state space $\ket{\psi_1}\oplus\ket{\psi_2}$
composed of two subspaces $\ket{\psi_\lambda}$ ($\lambda=1,2$).
The physics of each subspace $\ket{\psi_\lambda}$ is governed by the Hamiltonian $H_\lambda$.
$V$ describes the coupling between the two subspaces.
Generally, the subspace $\ket{\psi_\lambda}$ can be made up of several quantum states.
Hence $H_\lambda$ and $V$ can be cast into the matrix representation.
By choosing the basis $(\ket{\psi_1},\ket{\psi_2})^T$, the Hamiltonian of the system is written as
\begin{equation}
	H = \begin{pmatrix}
		H_1 & V \\ V^\dag & H_2
	\end{pmatrix}
	\equiv H_0 + U \,.
\end{equation}
where
\begin{equation}
	H_0 = \begin{pmatrix}
		H_1 \\ & H_2
	\end{pmatrix} \,,\quad
	U = \begin{pmatrix}
		& V \\ V^\dag
	\end{pmatrix} \,.
\end{equation}
Hereafter we regard the block-diagonal Hamiltonian $H_0$ as the nonperturbative term,
and the off-block-diagonal Hamiltonian $U$ as the perturbative one.
The Green's functions extracted from $H$ and $H_0$ are defined as (with respect to the variable $z$)
\begin{equation}
	G(z) = \frac{1}{z-H} \,,\quad
	G^{(0)}(z) = \frac{1}{z-H_0} \,. \label{eq:app:green-func-def}
\end{equation}
Since $H_0$ is block diagonal, its off-block-diagonal elements vanish.
It reveals that
\begin{equation}
	G_{12}^{(0)}=G_{21}^{(0)}=0 \,.
\end{equation}
Likewise for $U$, we have
\begin{equation}
	U_{11}=U_{22}=0 \,.
\end{equation}

We aim to obtain the effective Hamiltonian for the $\ket{\psi_1}$ subspace of interests,
yielding that the $\ket{\psi_2}$ subspace is adiabatically eliminated.
From the Dyson equation given as
\begin{equation}
	G = G^{(0)} +G^{(0)}UG \,.
\end{equation}
we only focus on the following block elements of the $G$ matrix,
\begin{align}
	G_{11} &= G_{11}^{(0)} + G_{11}^{(0)} U_{12} G_{21} \,,\\
	G_{21} &= G_{22}^{(0)} U_{21} G_{11} \,.
\end{align}
Submitting $G_{21}$ into $G_{11}$, we have
\begin{equation}
	G_{11} = G_{11}^{(0)} + G_{11}^{(0)} U_{12} G_{22}^{(0)} U_{21} G_{11} \,.
\end{equation}
The inverse of $G_{11}$ gives
\begin{align*}
	G_{11}^{-1} &= [G_{11}^{(0)}]^{-1}(1-G_{11}^{(0)} U_{12} G_{22}^{(0)} U_{21}) \\
	&= [G_{11}^{(0)}]^{-1}- U_{12} G_{22}^{(0)} U_{21} \,.
\end{align*}
Submitting Eq.(\ref{eq:app:green-func-def}), we get
\begin{equation*}
	z-H_{\rm eff} = z - H_1 - V\frac{1}{z-H_2}V^\dag \,.
\end{equation*}
By replacing $z$ by its first-order approximation $z\approx E_1$ with $E_1$ denoting the eigen-energy of $H_1$, we obtain the effective Hamiltonian for the $\ket{\psi_1}$ subspace written as
\begin{equation}
	H_{\rm eff} \approx H_1 + V\frac{1}{E_1-H_2}V^\dag \,. \label{eq:app:h-eff-final-form}
\end{equation}
We note that despite the fact that the above derivations are performed for obtaining a single-particle Hamiltonian,
the form of Eq.(\ref{eq:app:h-eff-final-form}) is still valid in the context of many-body systems with interactions.
This implies that the presence of interactions does not alter the structure of the engineered SOC.

%----------------------------------------------------------------------------------------
\vfill
\bibliographystyle{apsrev4-1}
\bibliography{reference}

\begin{thebibliography}{101}%
\makeatletter
\providecommand \@ifxundefined [1]{%
 \@ifx{#1\undefined}
}%
\providecommand \@ifnum [1]{%
 \ifnum #1\expandafter \@firstoftwo
 \else \expandafter \@secondoftwo
 \fi
}%
\providecommand \@ifx [1]{%
 \ifx #1\expandafter \@firstoftwo
 \else \expandafter \@secondoftwo
 \fi
}%
\providecommand \natexlab [1]{#1}%
\providecommand \enquote  [1]{``#1''}%
\providecommand \bibnamefont  [1]{#1}%
\providecommand \bibfnamefont [1]{#1}%
\providecommand \citenamefont [1]{#1}%
\providecommand \href@noop [0]{\@secondoftwo}%
\providecommand \href [0]{\begingroup \@sanitize@url \@href}%
\providecommand \@href[1]{\@@startlink{#1}\@@href}%
\providecommand \@@href[1]{\endgroup#1\@@endlink}%
\providecommand \@sanitize@url [0]{\catcode `\\12\catcode `\$12\catcode `\&12\catcode `\#12\catcode `\^12\catcode `\_12\catcode `\%12\relax}%
\providecommand \@@startlink[1]{}%
\providecommand \@@endlink[0]{}%
\providecommand \url  [0]{\begingroup\@sanitize@url \@url }%
\providecommand \@url [1]{\endgroup\@href {#1}{\urlprefix }}%
\providecommand \urlprefix  [0]{URL }%
\providecommand \Eprint [0]{\href }%
\providecommand \doibase [0]{http://dx.doi.org/}%
\providecommand \selectlanguage [0]{\@gobble}%
\providecommand \bibinfo  [0]{\@secondoftwo}%
\providecommand \bibfield  [0]{\@secondoftwo}%
\providecommand \translation [1]{[#1]}%
\providecommand \BibitemOpen [0]{}%
\providecommand \bibitemStop [0]{}%
\providecommand \bibitemNoStop [0]{.\EOS\space}%
\providecommand \EOS [0]{\spacefactor3000\relax}%
\providecommand \BibitemShut  [1]{\csname bibitem#1\endcsname}%
\let\auto@bib@innerbib\@empty
%</preamble>
\bibitem [{\citenamefont {Zhang}\ \emph {et~al.}(2018)\citenamefont {Zhang}, \citenamefont {Zhu}, \citenamefont {Zhao}, \citenamefont {Yan},\ and\ \citenamefont {Zhu}}]{Zhang2018Oct-topo-review}%
  \BibitemOpen
  \bibfield  {author} {\bibinfo {author} {\bibfnamefont {D.-W.}\ \bibnamefont {Zhang}}, \bibinfo {author} {\bibfnamefont {Y.-Q.}\ \bibnamefont {Zhu}}, \bibinfo {author} {\bibfnamefont {Y.~X.}\ \bibnamefont {Zhao}}, \bibinfo {author} {\bibfnamefont {H.}~\bibnamefont {Yan}}, \ and\ \bibinfo {author} {\bibfnamefont {S.-L.}\ \bibnamefont {Zhu}},\ }\href {\doibase 10.1080/00018732.2019.1594094} {\bibfield  {journal} {\bibinfo  {journal} {Adv. Phys.}\ }\textbf {\bibinfo {volume} {64}},\ \bibinfo {pages} {253} (\bibinfo {year} {2018})}\BibitemShut {NoStop}%
\bibitem [{\citenamefont {Dalibard}\ \emph {et~al.}(2011)\citenamefont {Dalibard}, \citenamefont {Gerbier}, \citenamefont {Juzeli{\ifmmode\bar{u}\else\={u}\fi}nas},\ and\ \citenamefont {{\ifmmode\ddot{O}\else\"{O}\fi}hberg}}]{Dalibard2011Nov-artificial-field-review}%
  \BibitemOpen
  \bibfield  {author} {\bibinfo {author} {\bibfnamefont {J.}~\bibnamefont {Dalibard}}, \bibinfo {author} {\bibfnamefont {F.}~\bibnamefont {Gerbier}}, \bibinfo {author} {\bibfnamefont {G.}~\bibnamefont {Juzeli{\ifmmode\bar{u}\else\={u}\fi}nas}}, \ and\ \bibinfo {author} {\bibfnamefont {P.}~\bibnamefont {{\ifmmode\ddot{O}\else\"{O}\fi}hberg}},\ }\href {\doibase 10.1103/RevModPhys.83.1523} {\bibfield  {journal} {\bibinfo  {journal} {Rev. Mod. Phys.}\ }\textbf {\bibinfo {volume} {83}},\ \bibinfo {pages} {1523} (\bibinfo {year} {2011})}\BibitemShut {NoStop}%
\bibitem [{\citenamefont {Bloch}\ \emph {et~al.}(2012)\citenamefont {Bloch}, \citenamefont {Dalibard},\ and\ \citenamefont {Nascimb{\ifmmode\grave{e}\else\`{e}\fi}ne}}]{Bloch2012Apr-simu-review}%
  \BibitemOpen
  \bibfield  {author} {\bibinfo {author} {\bibfnamefont {I.}~\bibnamefont {Bloch}}, \bibinfo {author} {\bibfnamefont {J.}~\bibnamefont {Dalibard}}, \ and\ \bibinfo {author} {\bibfnamefont {S.}~\bibnamefont {Nascimb{\ifmmode\grave{e}\else\`{e}\fi}ne}},\ }\href {\doibase 10.1038/nphys2259} {\bibfield  {journal} {\bibinfo  {journal} {Nat. Phys.}\ }\textbf {\bibinfo {volume} {8}},\ \bibinfo {pages} {267} (\bibinfo {year} {2012})}\BibitemShut {NoStop}%
\bibitem [{\citenamefont {Gross}\ and\ \citenamefont {Bloch}(2017)}]{Gross2017Sep-simu-review}%
  \BibitemOpen
  \bibfield  {author} {\bibinfo {author} {\bibfnamefont {C.}~\bibnamefont {Gross}}\ and\ \bibinfo {author} {\bibfnamefont {I.}~\bibnamefont {Bloch}},\ }\href {\doibase 10.1126/science.aal3837} {\bibfield  {journal} {\bibinfo  {journal} {Science}\ }\textbf {\bibinfo {volume} {357}},\ \bibinfo {pages} {995} (\bibinfo {year} {2017})}\BibitemShut {NoStop}%
\bibitem [{\citenamefont {Goldman}\ \emph {et~al.}(2014)\citenamefont {Goldman}, \citenamefont {Juzeli{\ifmmode\bar{u}\else\={u}\fi}nas}, \citenamefont {{\ifmmode\ddot{O}\else\"{O}\fi}hberg},\ and\ \citenamefont {Spielman}}]{Goldman2014Nov-artificial-field-review}%
  \BibitemOpen
  \bibfield  {author} {\bibinfo {author} {\bibfnamefont {N.}~\bibnamefont {Goldman}}, \bibinfo {author} {\bibfnamefont {G.}~\bibnamefont {Juzeli{\ifmmode\bar{u}\else\={u}\fi}nas}}, \bibinfo {author} {\bibfnamefont {P.}~\bibnamefont {{\ifmmode\ddot{O}\else\"{O}\fi}hberg}}, \ and\ \bibinfo {author} {\bibfnamefont {I.~B.}\ \bibnamefont {Spielman}},\ }\href {\doibase 10.1088/0034-4885/77/12/126401} {\bibfield  {journal} {\bibinfo  {journal} {Rep. Prog. Phys.}\ }\textbf {\bibinfo {volume} {77}},\ \bibinfo {pages} {126401} (\bibinfo {year} {2014})}\BibitemShut {NoStop}%
\bibitem [{\citenamefont {Eckardt}(2017)}]{Eckardt2017Mar-artificial-field-review}%
  \BibitemOpen
  \bibfield  {author} {\bibinfo {author} {\bibfnamefont {A.}~\bibnamefont {Eckardt}},\ }\href {\doibase 10.1103/RevModPhys.89.011004} {\bibfield  {journal} {\bibinfo  {journal} {Rev. Mod. Phys.}\ }\textbf {\bibinfo {volume} {89}},\ \bibinfo {pages} {011004} (\bibinfo {year} {2017})}\BibitemShut {NoStop}%
\bibitem [{\citenamefont {Hasan}\ and\ \citenamefont {Kane}(2010)}]{Hasan2010Nov-topo-review}%
  \BibitemOpen
  \bibfield  {author} {\bibinfo {author} {\bibfnamefont {M.~Z.}\ \bibnamefont {Hasan}}\ and\ \bibinfo {author} {\bibfnamefont {C.~L.}\ \bibnamefont {Kane}},\ }\href {\doibase 10.1103/RevModPhys.82.3045} {\bibfield  {journal} {\bibinfo  {journal} {Rev. Mod. Phys.}\ }\textbf {\bibinfo {volume} {82}},\ \bibinfo {pages} {3045} (\bibinfo {year} {2010})}\BibitemShut {NoStop}%
\bibitem [{\citenamefont {Qi}\ and\ \citenamefont {Zhang}(2011)}]{Qi2011Oct-topo-review}%
  \BibitemOpen
  \bibfield  {author} {\bibinfo {author} {\bibfnamefont {X.-L.}\ \bibnamefont {Qi}}\ and\ \bibinfo {author} {\bibfnamefont {S.-C.}\ \bibnamefont {Zhang}},\ }\href {\doibase 10.1103/RevModPhys.83.1057} {\bibfield  {journal} {\bibinfo  {journal} {Rev. Mod. Phys.}\ }\textbf {\bibinfo {volume} {83}},\ \bibinfo {pages} {1057} (\bibinfo {year} {2011})}\BibitemShut {NoStop}%
\bibitem [{\citenamefont {Beeler}\ \emph {et~al.}(2013)\citenamefont {Beeler}, \citenamefont {Williams}, \citenamefont {Jiménez-García}, \citenamefont {LeBlanc}, \citenamefont {Perry},\ and\ \citenamefont {Spielman}}]{Beeler2013}%
  \BibitemOpen
  \bibfield  {author} {\bibinfo {author} {\bibfnamefont {M.~C.}\ \bibnamefont {Beeler}}, \bibinfo {author} {\bibfnamefont {R.~A.}\ \bibnamefont {Williams}}, \bibinfo {author} {\bibfnamefont {K.}~\bibnamefont {Jiménez-García}}, \bibinfo {author} {\bibfnamefont {L.~J.}\ \bibnamefont {LeBlanc}}, \bibinfo {author} {\bibfnamefont {A.~R.}\ \bibnamefont {Perry}}, \ and\ \bibinfo {author} {\bibfnamefont {I.~B.}\ \bibnamefont {Spielman}},\ }\href {\doibase 10.1038/nature12185} {\bibfield  {journal} {\bibinfo  {journal} {Nature}\ }\textbf {\bibinfo {volume} {498}},\ \bibinfo {pages} {201} (\bibinfo {year} {2013})}\BibitemShut {NoStop}%
\bibitem [{\citenamefont {Zhu}\ \emph {et~al.}(2006)\citenamefont {Zhu}, \citenamefont {Fu}, \citenamefont {Wu}, \citenamefont {Zhang},\ and\ \citenamefont {Duan}}]{Zhu2006Dec-soc-appli}%
  \BibitemOpen
  \bibfield  {author} {\bibinfo {author} {\bibfnamefont {S.-L.}\ \bibnamefont {Zhu}}, \bibinfo {author} {\bibfnamefont {H.}~\bibnamefont {Fu}}, \bibinfo {author} {\bibfnamefont {C.-J.}\ \bibnamefont {Wu}}, \bibinfo {author} {\bibfnamefont {S.-C.}\ \bibnamefont {Zhang}}, \ and\ \bibinfo {author} {\bibfnamefont {L.-M.}\ \bibnamefont {Duan}},\ }\href {\doibase 10.1103/PhysRevLett.97.240401} {\bibfield  {journal} {\bibinfo  {journal} {Phys. Rev. Lett.}\ }\textbf {\bibinfo {volume} {97}},\ \bibinfo {pages} {240401} (\bibinfo {year} {2006})}\BibitemShut {NoStop}%
\bibitem [{\citenamefont {Liu}\ \emph {et~al.}(2007)\citenamefont {Liu}, \citenamefont {Liu}, \citenamefont {Kwek},\ and\ \citenamefont {Oh}}]{XJLiu2007}%
  \BibitemOpen
  \bibfield  {author} {\bibinfo {author} {\bibfnamefont {X.-J.}\ \bibnamefont {Liu}}, \bibinfo {author} {\bibfnamefont {X.}~\bibnamefont {Liu}}, \bibinfo {author} {\bibfnamefont {L.~C.}\ \bibnamefont {Kwek}}, \ and\ \bibinfo {author} {\bibfnamefont {C.~H.}\ \bibnamefont {Oh}},\ }\href {\doibase 10.1103/physrevlett.98.026602} {\bibfield  {journal} {\bibinfo  {journal} {Phys. Rev. Lett.}\ }\textbf {\bibinfo {volume} {98}},\ \bibinfo {pages} {026602} (\bibinfo {year} {2007})}\BibitemShut {NoStop}%
\bibitem [{\citenamefont {Kennedy}\ \emph {et~al.}(2013)\citenamefont {Kennedy}, \citenamefont {Siviloglou}, \citenamefont {Miyake}, \citenamefont {Burton},\ and\ \citenamefont {Ketterle}}]{Kennedy2013}%
  \BibitemOpen
  \bibfield  {author} {\bibinfo {author} {\bibfnamefont {C.~J.}\ \bibnamefont {Kennedy}}, \bibinfo {author} {\bibfnamefont {G.~A.}\ \bibnamefont {Siviloglou}}, \bibinfo {author} {\bibfnamefont {H.}~\bibnamefont {Miyake}}, \bibinfo {author} {\bibfnamefont {W.~C.}\ \bibnamefont {Burton}}, \ and\ \bibinfo {author} {\bibfnamefont {W.}~\bibnamefont {Ketterle}},\ }\href {\doibase 10.1103/physrevlett.111.225301} {\bibfield  {journal} {\bibinfo  {journal} {Phys. Rev. Lett.}\ }\textbf {\bibinfo {volume} {111}},\ \bibinfo {pages} {225301} (\bibinfo {year} {2013})}\BibitemShut {NoStop}%
\bibitem [{\citenamefont {Lv}\ \emph {et~al.}(2021)\citenamefont {Lv}, \citenamefont {Du}, \citenamefont {Liang}, \citenamefont {Liu}, \citenamefont {Liang}, \citenamefont {Chen}, \citenamefont {Zhou}, \citenamefont {Zhang}, \citenamefont {Zhang}, \citenamefont {Ai}, \citenamefont {Yan},\ and\ \citenamefont {Zhu}}]{Lv2021Sep-soc-exp-Rb87}%
  \BibitemOpen
  \bibfield  {author} {\bibinfo {author} {\bibfnamefont {Q.-X.}\ \bibnamefont {Lv}}, \bibinfo {author} {\bibfnamefont {Y.-X.}\ \bibnamefont {Du}}, \bibinfo {author} {\bibfnamefont {Z.-T.}\ \bibnamefont {Liang}}, \bibinfo {author} {\bibfnamefont {H.-Z.}\ \bibnamefont {Liu}}, \bibinfo {author} {\bibfnamefont {J.-H.}\ \bibnamefont {Liang}}, \bibinfo {author} {\bibfnamefont {L.-Q.}\ \bibnamefont {Chen}}, \bibinfo {author} {\bibfnamefont {L.-M.}\ \bibnamefont {Zhou}}, \bibinfo {author} {\bibfnamefont {S.-C.}\ \bibnamefont {Zhang}}, \bibinfo {author} {\bibfnamefont {D.-W.}\ \bibnamefont {Zhang}}, \bibinfo {author} {\bibfnamefont {B.-Q.}\ \bibnamefont {Ai}}, \bibinfo {author} {\bibfnamefont {H.}~\bibnamefont {Yan}}, \ and\ \bibinfo {author} {\bibfnamefont {S.-L.}\ \bibnamefont {Zhu}},\ }\href {\doibase 10.1103/PhysRevLett.127.136802} {\bibfield  {journal} {\bibinfo  {journal} {Phys. Rev. Lett.}\ }\textbf {\bibinfo {volume} {127}},\ \bibinfo {pages} {136802} (\bibinfo {year} {2021})}\BibitemShut {NoStop}%
\bibitem [{\citenamefont {Jotzu}\ \emph {et~al.}(2014)\citenamefont {Jotzu}, \citenamefont {Messer}, \citenamefont {Desbuquois}, \citenamefont {Lebrat}, \citenamefont {Uehlinger}, \citenamefont {Greif},\ and\ \citenamefont {Esslinger}}]{Jotzu2014}%
  \BibitemOpen
  \bibfield  {author} {\bibinfo {author} {\bibfnamefont {G.}~\bibnamefont {Jotzu}}, \bibinfo {author} {\bibfnamefont {M.}~\bibnamefont {Messer}}, \bibinfo {author} {\bibfnamefont {R.}~\bibnamefont {Desbuquois}}, \bibinfo {author} {\bibfnamefont {M.}~\bibnamefont {Lebrat}}, \bibinfo {author} {\bibfnamefont {T.}~\bibnamefont {Uehlinger}}, \bibinfo {author} {\bibfnamefont {D.}~\bibnamefont {Greif}}, \ and\ \bibinfo {author} {\bibfnamefont {T.}~\bibnamefont {Esslinger}},\ }\href {\doibase 10.1038/nature13915} {\bibfield  {journal} {\bibinfo  {journal} {Nature}\ }\textbf {\bibinfo {volume} {515}},\ \bibinfo {pages} {237} (\bibinfo {year} {2014})}\BibitemShut {NoStop}%
\bibitem [{\citenamefont {Shao}\ \emph {et~al.}(2008)\citenamefont {Shao}, \citenamefont {Zhu}, \citenamefont {Sheng}, \citenamefont {Xing},\ and\ \citenamefont {Wang}}]{LBShao2008}%
  \BibitemOpen
  \bibfield  {author} {\bibinfo {author} {\bibfnamefont {L.~B.}\ \bibnamefont {Shao}}, \bibinfo {author} {\bibfnamefont {S.-L.}\ \bibnamefont {Zhu}}, \bibinfo {author} {\bibfnamefont {L.}~\bibnamefont {Sheng}}, \bibinfo {author} {\bibfnamefont {D.~Y.}\ \bibnamefont {Xing}}, \ and\ \bibinfo {author} {\bibfnamefont {Z.~D.}\ \bibnamefont {Wang}},\ }\href {\doibase 10.1103/physrevlett.101.246810} {\bibfield  {journal} {\bibinfo  {journal} {Phys. Rev. Lett.}\ }\textbf {\bibinfo {volume} {101}},\ \bibinfo {pages} {246810} (\bibinfo {year} {2008})}\BibitemShut {NoStop}%
\bibitem [{\citenamefont {Wu}(2008)}]{CJWu2008}%
  \BibitemOpen
  \bibfield  {author} {\bibinfo {author} {\bibfnamefont {C.}~\bibnamefont {Wu}},\ }\href {\doibase 10.1103/physrevlett.101.186807} {\bibfield  {journal} {\bibinfo  {journal} {Phys. Rev. Lett.}\ }\textbf {\bibinfo {volume} {101}},\ \bibinfo {pages} {186807} (\bibinfo {year} {2008})}\BibitemShut {NoStop}%
\bibitem [{\citenamefont {Liu}\ \emph {et~al.}(2018)\citenamefont {Liu}, \citenamefont {Shao}, \citenamefont {Hou},\ and\ \citenamefont {Xue}}]{SLiu2018}%
  \BibitemOpen
  \bibfield  {author} {\bibinfo {author} {\bibfnamefont {S.}~\bibnamefont {Liu}}, \bibinfo {author} {\bibfnamefont {L.~B.}\ \bibnamefont {Shao}}, \bibinfo {author} {\bibfnamefont {Q.-Z.}\ \bibnamefont {Hou}}, \ and\ \bibinfo {author} {\bibfnamefont {Z.-Y.}\ \bibnamefont {Xue}},\ }\href {\doibase 10.1088/1361-648x/aaab89} {\bibfield  {journal} {\bibinfo  {journal} {J. Phys.-Condes. Matter}\ }\textbf {\bibinfo {volume} {30}},\ \bibinfo {pages} {124001} (\bibinfo {year} {2018})}\BibitemShut {NoStop}%
\bibitem [{\citenamefont {Vasić}\ \emph {et~al.}(2015)\citenamefont {Vasić}, \citenamefont {Petrescu}, \citenamefont {Le~Hur},\ and\ \citenamefont {Hofstetter}}]{Vasic2015}%
  \BibitemOpen
  \bibfield  {author} {\bibinfo {author} {\bibfnamefont {I.}~\bibnamefont {Vasić}}, \bibinfo {author} {\bibfnamefont {A.}~\bibnamefont {Petrescu}}, \bibinfo {author} {\bibfnamefont {K.}~\bibnamefont {Le~Hur}}, \ and\ \bibinfo {author} {\bibfnamefont {W.}~\bibnamefont {Hofstetter}},\ }\href {\doibase 10.1103/physrevb.91.094502} {\bibfield  {journal} {\bibinfo  {journal} {Phys. Rev. B}\ }\textbf {\bibinfo {volume} {91}},\ \bibinfo {pages} {094502} (\bibinfo {year} {2015})}\BibitemShut {NoStop}%
\bibitem [{\citenamefont {Deng}\ \emph {et~al.}(2014{\natexlab{a}})\citenamefont {Deng}, \citenamefont {Wang},\ and\ \citenamefont {Duan}}]{DLDeng2014}%
  \BibitemOpen
  \bibfield  {author} {\bibinfo {author} {\bibfnamefont {D.-L.}\ \bibnamefont {Deng}}, \bibinfo {author} {\bibfnamefont {S.-T.}\ \bibnamefont {Wang}}, \ and\ \bibinfo {author} {\bibfnamefont {L.-M.}\ \bibnamefont {Duan}},\ }\href {\doibase 10.1103/physreva.90.041601} {\bibfield  {journal} {\bibinfo  {journal} {Physical Review A}\ }\textbf {\bibinfo {volume} {90}},\ \bibinfo {pages} {041601} (\bibinfo {year} {2014}{\natexlab{a}})}\BibitemShut {NoStop}%
\bibitem [{\citenamefont {Anisimovas}\ \emph {et~al.}(2014)\citenamefont {Anisimovas}, \citenamefont {Gerbier}, \citenamefont {Andrijauskas},\ and\ \citenamefont {Goldman}}]{Anisimovas2014}%
  \BibitemOpen
  \bibfield  {author} {\bibinfo {author} {\bibfnamefont {E.}~\bibnamefont {Anisimovas}}, \bibinfo {author} {\bibfnamefont {F.}~\bibnamefont {Gerbier}}, \bibinfo {author} {\bibfnamefont {T.}~\bibnamefont {Andrijauskas}}, \ and\ \bibinfo {author} {\bibfnamefont {N.}~\bibnamefont {Goldman}},\ }\href {\doibase 10.1103/physreva.89.013632} {\bibfield  {journal} {\bibinfo  {journal} {Physical Review A}\ }\textbf {\bibinfo {volume} {89}},\ \bibinfo {pages} {013632} (\bibinfo {year} {2014})}\BibitemShut {NoStop}%
\bibitem [{\citenamefont {Zhu}\ \emph {et~al.}(2007)\citenamefont {Zhu}, \citenamefont {Wang},\ and\ \citenamefont {Duan}}]{SLZhu2007}%
  \BibitemOpen
  \bibfield  {author} {\bibinfo {author} {\bibfnamefont {S.-L.}\ \bibnamefont {Zhu}}, \bibinfo {author} {\bibfnamefont {B.}~\bibnamefont {Wang}}, \ and\ \bibinfo {author} {\bibfnamefont {L.-M.}\ \bibnamefont {Duan}},\ }\href {\doibase 10.1103/physrevlett.98.260402} {\bibfield  {journal} {\bibinfo  {journal} {Phys. Rev. Lett.}\ }\textbf {\bibinfo {volume} {98}},\ \bibinfo {pages} {260402} (\bibinfo {year} {2007})}\BibitemShut {NoStop}%
\bibitem [{\citenamefont {Tarruell}\ \emph {et~al.}(2012)\citenamefont {Tarruell}, \citenamefont {Greif}, \citenamefont {Uehlinger}, \citenamefont {Jotzu},\ and\ \citenamefont {Esslinger}}]{Tarruell2012}%
  \BibitemOpen
  \bibfield  {author} {\bibinfo {author} {\bibfnamefont {L.}~\bibnamefont {Tarruell}}, \bibinfo {author} {\bibfnamefont {D.}~\bibnamefont {Greif}}, \bibinfo {author} {\bibfnamefont {T.}~\bibnamefont {Uehlinger}}, \bibinfo {author} {\bibfnamefont {G.}~\bibnamefont {Jotzu}}, \ and\ \bibinfo {author} {\bibfnamefont {T.}~\bibnamefont {Esslinger}},\ }\href {\doibase 10.1038/nature10871} {\bibfield  {journal} {\bibinfo  {journal} {Nature}\ }\textbf {\bibinfo {volume} {483}},\ \bibinfo {pages} {302} (\bibinfo {year} {2012})}\BibitemShut {NoStop}%
\bibitem [{\citenamefont {Gomes}\ \emph {et~al.}(2012)\citenamefont {Gomes}, \citenamefont {Mar}, \citenamefont {Ko}, \citenamefont {Guinea},\ and\ \citenamefont {Manoharan}}]{Gomes2012}%
  \BibitemOpen
  \bibfield  {author} {\bibinfo {author} {\bibfnamefont {K.~K.}\ \bibnamefont {Gomes}}, \bibinfo {author} {\bibfnamefont {W.}~\bibnamefont {Mar}}, \bibinfo {author} {\bibfnamefont {W.}~\bibnamefont {Ko}}, \bibinfo {author} {\bibfnamefont {F.}~\bibnamefont {Guinea}}, \ and\ \bibinfo {author} {\bibfnamefont {H.~C.}\ \bibnamefont {Manoharan}},\ }\href {\doibase 10.1038/nature10941} {\bibfield  {journal} {\bibinfo  {journal} {Nature}\ }\textbf {\bibinfo {volume} {483}},\ \bibinfo {pages} {306} (\bibinfo {year} {2012})}\BibitemShut {NoStop}%
\bibitem [{\citenamefont {Duca}\ \emph {et~al.}(2015)\citenamefont {Duca}, \citenamefont {Li}, \citenamefont {Reitter}, \citenamefont {Bloch}, \citenamefont {Schleier-Smith},\ and\ \citenamefont {Schneider}}]{Duca2015}%
  \BibitemOpen
  \bibfield  {author} {\bibinfo {author} {\bibfnamefont {L.}~\bibnamefont {Duca}}, \bibinfo {author} {\bibfnamefont {T.}~\bibnamefont {Li}}, \bibinfo {author} {\bibfnamefont {M.}~\bibnamefont {Reitter}}, \bibinfo {author} {\bibfnamefont {I.}~\bibnamefont {Bloch}}, \bibinfo {author} {\bibfnamefont {M.}~\bibnamefont {Schleier-Smith}}, \ and\ \bibinfo {author} {\bibfnamefont {U.}~\bibnamefont {Schneider}},\ }\href {\doibase 10.1126/science.1259052} {\bibfield  {journal} {\bibinfo  {journal} {Science}\ }\textbf {\bibinfo {volume} {347}},\ \bibinfo {pages} {288} (\bibinfo {year} {2015})}\BibitemShut {NoStop}%
\bibitem [{\citenamefont {Zhang}\ \emph {et~al.}(2015)\citenamefont {Zhang}, \citenamefont {Zhu},\ and\ \citenamefont {Wang}}]{Zhang2015Jul-soc-appli}%
  \BibitemOpen
  \bibfield  {author} {\bibinfo {author} {\bibfnamefont {D.-W.}\ \bibnamefont {Zhang}}, \bibinfo {author} {\bibfnamefont {S.-L.}\ \bibnamefont {Zhu}}, \ and\ \bibinfo {author} {\bibfnamefont {Z.~D.}\ \bibnamefont {Wang}},\ }\href {\doibase 10.1103/PhysRevA.92.013632} {\bibfield  {journal} {\bibinfo  {journal} {Phys. Rev. A}\ }\textbf {\bibinfo {volume} {92}},\ \bibinfo {pages} {013632} (\bibinfo {year} {2015})}\BibitemShut {NoStop}%
\bibitem [{\citenamefont {Zhu}\ \emph {et~al.}(2011)\citenamefont {Zhu}, \citenamefont {Shao}, \citenamefont {Wang},\ and\ \citenamefont {Duan}}]{SLZhu2011}%
  \BibitemOpen
  \bibfield  {author} {\bibinfo {author} {\bibfnamefont {S.-L.}\ \bibnamefont {Zhu}}, \bibinfo {author} {\bibfnamefont {L.-B.}\ \bibnamefont {Shao}}, \bibinfo {author} {\bibfnamefont {Z.~D.}\ \bibnamefont {Wang}}, \ and\ \bibinfo {author} {\bibfnamefont {L.-M.}\ \bibnamefont {Duan}},\ }\href {\doibase 10.1103/physrevlett.106.100404} {\bibfield  {journal} {\bibinfo  {journal} {Phys. Rev. Lett.}\ }\textbf {\bibinfo {volume} {106}},\ \bibinfo {pages} {100404} (\bibinfo {year} {2011})}\BibitemShut {NoStop}%
\bibitem [{\citenamefont {Qu}\ \emph {et~al.}(2015)\citenamefont {Qu}, \citenamefont {Gong}, \citenamefont {Xu}, \citenamefont {Tewari},\ and\ \citenamefont {Zhang}}]{CLQu2015}%
  \BibitemOpen
  \bibfield  {author} {\bibinfo {author} {\bibfnamefont {C.}~\bibnamefont {Qu}}, \bibinfo {author} {\bibfnamefont {M.}~\bibnamefont {Gong}}, \bibinfo {author} {\bibfnamefont {Y.}~\bibnamefont {Xu}}, \bibinfo {author} {\bibfnamefont {S.}~\bibnamefont {Tewari}}, \ and\ \bibinfo {author} {\bibfnamefont {C.}~\bibnamefont {Zhang}},\ }\href {\doibase 10.1103/physreva.92.023621} {\bibfield  {journal} {\bibinfo  {journal} {Physical Review A}\ }\textbf {\bibinfo {volume} {92}},\ \bibinfo {pages} {023621} (\bibinfo {year} {2015})}\BibitemShut {NoStop}%
\bibitem [{\citenamefont {Guo}\ \emph {et~al.}(2019)\citenamefont {Guo}, \citenamefont {Lin}, \citenamefont {Zhao}, \citenamefont {Lou},\ and\ \citenamefont {Chen}}]{YWGuo2019}%
  \BibitemOpen
  \bibfield  {author} {\bibinfo {author} {\bibfnamefont {Y.}~\bibnamefont {Guo}}, \bibinfo {author} {\bibfnamefont {Z.}~\bibnamefont {Lin}}, \bibinfo {author} {\bibfnamefont {J.-Q.}\ \bibnamefont {Zhao}}, \bibinfo {author} {\bibfnamefont {J.}~\bibnamefont {Lou}}, \ and\ \bibinfo {author} {\bibfnamefont {Y.}~\bibnamefont {Chen}},\ }\href {\doibase 10.1038/s41598-019-54670-5} {\bibfield  {journal} {\bibinfo  {journal} {Scientific Reports}\ }\textbf {\bibinfo {volume} {9}},\ \bibinfo {pages} {18516} (\bibinfo {year} {2019})}\BibitemShut {NoStop}%
\bibitem [{\citenamefont {Li}\ \emph {et~al.}(2022)\citenamefont {Li}, \citenamefont {Zou}, \citenamefont {Du}, \citenamefont {Lv}, \citenamefont {Huang}, \citenamefont {Liang}, \citenamefont {Zhang}, \citenamefont {Yan}, \citenamefont {Zhang},\ and\ \citenamefont {Zhu}}]{JZLi2022}%
  \BibitemOpen
  \bibfield  {author} {\bibinfo {author} {\bibfnamefont {J.-Z.}\ \bibnamefont {Li}}, \bibinfo {author} {\bibfnamefont {C.-J.}\ \bibnamefont {Zou}}, \bibinfo {author} {\bibfnamefont {Y.-X.}\ \bibnamefont {Du}}, \bibinfo {author} {\bibfnamefont {Q.-X.}\ \bibnamefont {Lv}}, \bibinfo {author} {\bibfnamefont {W.}~\bibnamefont {Huang}}, \bibinfo {author} {\bibfnamefont {Z.-T.}\ \bibnamefont {Liang}}, \bibinfo {author} {\bibfnamefont {D.-W.}\ \bibnamefont {Zhang}}, \bibinfo {author} {\bibfnamefont {H.}~\bibnamefont {Yan}}, \bibinfo {author} {\bibfnamefont {S.}~\bibnamefont {Zhang}}, \ and\ \bibinfo {author} {\bibfnamefont {S.-L.}\ \bibnamefont {Zhu}},\ }\href {\doibase 10.1103/physrevlett.129.220402} {\bibfield  {journal} {\bibinfo  {journal} {Phys. Rev. Lett.}\ }\textbf {\bibinfo {volume} {129}},\ \bibinfo {pages} {220402} (\bibinfo {year} {2022})}\BibitemShut {NoStop}%
\bibitem [{\citenamefont {Galitski}\ and\ \citenamefont {Spielman}(2013)}]{Galitski2013-soc-review}%
  \BibitemOpen
  \bibfield  {author} {\bibinfo {author} {\bibfnamefont {V.}~\bibnamefont {Galitski}}\ and\ \bibinfo {author} {\bibfnamefont {I.~B.}\ \bibnamefont {Spielman}},\ }\href {\doibase 10.1038/nature11841} {\bibfield  {journal} {\bibinfo  {journal} {Nature}\ }\textbf {\bibinfo {volume} {494}},\ \bibinfo {pages} {49} (\bibinfo {year} {2013})}\BibitemShut {NoStop}%
\bibitem [{\citenamefont {Zhang}\ \emph {et~al.}(2013)\citenamefont {Zhang}, \citenamefont {Hu}, \citenamefont {Liu},\ and\ \citenamefont {Pu}}]{Zhang2013Dec-soc-review}%
  \BibitemOpen
  \bibfield  {author} {\bibinfo {author} {\bibfnamefont {J.}~\bibnamefont {Zhang}}, \bibinfo {author} {\bibfnamefont {H.}~\bibnamefont {Hu}}, \bibinfo {author} {\bibfnamefont {X.-J.}\ \bibnamefont {Liu}}, \ and\ \bibinfo {author} {\bibfnamefont {H.}~\bibnamefont {Pu}},\ }in\ \href {\doibase 10.1142/9789814590174_0002} {\emph {\bibinfo {booktitle} {{Annual Review of Cold Atoms and Molecules}}}},\ Vol.~\bibinfo {volume} {2}\ (\bibinfo  {publisher} {World Scientific},\ \bibinfo {address} {Singapore},\ \bibinfo {year} {2013})\ pp.\ \bibinfo {pages} {81--143}\BibitemShut {NoStop}%
\bibitem [{\citenamefont {Zhai}(2015)}]{Zhai2015Feb-soc-review}%
  \BibitemOpen
  \bibfield  {author} {\bibinfo {author} {\bibfnamefont {H.}~\bibnamefont {Zhai}},\ }\href {\doibase 10.1088/0034-4885/78/2/026001} {\bibfield  {journal} {\bibinfo  {journal} {Rep. Prog. Phys.}\ }\textbf {\bibinfo {volume} {78}},\ \bibinfo {pages} {026001} (\bibinfo {year} {2015})}\BibitemShut {NoStop}%
\bibitem [{\citenamefont {Zhang}\ and\ \citenamefont {Jo}(2019)}]{Zhang2019May-soc-review}%
  \BibitemOpen
  \bibfield  {author} {\bibinfo {author} {\bibfnamefont {S.}~\bibnamefont {Zhang}}\ and\ \bibinfo {author} {\bibfnamefont {G.-B.}\ \bibnamefont {Jo}},\ }\href {\doibase 10.1016/j.jpcs.2018.04.033} {\bibfield  {journal} {\bibinfo  {journal} {J. Phys. Chem. Solids}\ }\textbf {\bibinfo {volume} {128}},\ \bibinfo {pages} {75} (\bibinfo {year} {2019})}\BibitemShut {NoStop}%
\bibitem [{\citenamefont {Osterloh}\ \emph {et~al.}(2005)\citenamefont {Osterloh}, \citenamefont {Baig}, \citenamefont {Santos}, \citenamefont {Zoller},\ and\ \citenamefont {Lewenstein}}]{Osterloh2005Jun-soc-theo}%
  \BibitemOpen
  \bibfield  {author} {\bibinfo {author} {\bibfnamefont {K.}~\bibnamefont {Osterloh}}, \bibinfo {author} {\bibfnamefont {M.}~\bibnamefont {Baig}}, \bibinfo {author} {\bibfnamefont {L.}~\bibnamefont {Santos}}, \bibinfo {author} {\bibfnamefont {P.}~\bibnamefont {Zoller}}, \ and\ \bibinfo {author} {\bibfnamefont {M.}~\bibnamefont {Lewenstein}},\ }\href {\doibase 10.1103/PhysRevLett.95.010403} {\bibfield  {journal} {\bibinfo  {journal} {Phys. Rev. Lett.}\ }\textbf {\bibinfo {volume} {95}},\ \bibinfo {pages} {010403} (\bibinfo {year} {2005})}\BibitemShut {NoStop}%
\bibitem [{\citenamefont {Ruseckas}\ \emph {et~al.}(2005)\citenamefont {Ruseckas}, \citenamefont {Juzeli{\ifmmode\bar{u}\else\={u}\fi}nas}, \citenamefont {{\ifmmode\ddot{O}\else\"{O}\fi}hberg},\ and\ \citenamefont {Fleischhauer}}]{Ruseckas2005Jun-soc-theo}%
  \BibitemOpen
  \bibfield  {author} {\bibinfo {author} {\bibfnamefont {J.}~\bibnamefont {Ruseckas}}, \bibinfo {author} {\bibfnamefont {G.}~\bibnamefont {Juzeli{\ifmmode\bar{u}\else\={u}\fi}nas}}, \bibinfo {author} {\bibfnamefont {P.}~\bibnamefont {{\ifmmode\ddot{O}\else\"{O}\fi}hberg}}, \ and\ \bibinfo {author} {\bibfnamefont {M.}~\bibnamefont {Fleischhauer}},\ }\href {\doibase 10.1103/PhysRevLett.95.010404} {\bibfield  {journal} {\bibinfo  {journal} {Phys. Rev. Lett.}\ }\textbf {\bibinfo {volume} {95}},\ \bibinfo {pages} {010404} (\bibinfo {year} {2005})}\BibitemShut {NoStop}%
\bibitem [{\citenamefont {Lin}\ \emph {et~al.}(2011)\citenamefont {Lin}, \citenamefont {Jim{\ifmmode\acute{e}\else\'{e}\fi}nez-Garc{\ifmmode\acute{\imath}\else\'{\i}\fi}a},\ and\ \citenamefont {Spielman}}]{Lin2011Mar-soc-exp-Rb87}%
  \BibitemOpen
  \bibfield  {author} {\bibinfo {author} {\bibfnamefont {Y.-J.}\ \bibnamefont {Lin}}, \bibinfo {author} {\bibfnamefont {K.}~\bibnamefont {Jim{\ifmmode\acute{e}\else\'{e}\fi}nez-Garc{\ifmmode\acute{\imath}\else\'{\i}\fi}a}}, \ and\ \bibinfo {author} {\bibfnamefont {I.~B.}\ \bibnamefont {Spielman}},\ }\href {\doibase 10.1038/nature09887} {\bibfield  {journal} {\bibinfo  {journal} {Nature}\ }\textbf {\bibinfo {volume} {471}},\ \bibinfo {pages} {83} (\bibinfo {year} {2011})}\BibitemShut {NoStop}%
\bibitem [{\citenamefont {Hamner}\ \emph {et~al.}(2014)\citenamefont {Hamner}, \citenamefont {Qu}, \citenamefont {Zhang}, \citenamefont {Chang}, \citenamefont {Gong}, \citenamefont {Zhang},\ and\ \citenamefont {Engels}}]{Hamner2014Jun-soc-exp-Rb87}%
  \BibitemOpen
  \bibfield  {author} {\bibinfo {author} {\bibfnamefont {C.}~\bibnamefont {Hamner}}, \bibinfo {author} {\bibfnamefont {C.}~\bibnamefont {Qu}}, \bibinfo {author} {\bibfnamefont {Y.}~\bibnamefont {Zhang}}, \bibinfo {author} {\bibfnamefont {J.}~\bibnamefont {Chang}}, \bibinfo {author} {\bibfnamefont {M.}~\bibnamefont {Gong}}, \bibinfo {author} {\bibfnamefont {C.}~\bibnamefont {Zhang}}, \ and\ \bibinfo {author} {\bibfnamefont {P.}~\bibnamefont {Engels}},\ }\href {\doibase 10.1038/ncomms5023} {\bibfield  {journal} {\bibinfo  {journal} {Nat. Commun.}\ }\textbf {\bibinfo {volume} {5}},\ \bibinfo {pages} {4023} (\bibinfo {year} {2014})}\BibitemShut {NoStop}%
\bibitem [{\citenamefont {Olson}\ \emph {et~al.}(2014)\citenamefont {Olson}, \citenamefont {Wang}, \citenamefont {Niffenegger}, \citenamefont {Li}, \citenamefont {Greene},\ and\ \citenamefont {Chen}}]{Olson2014Jul-soc-exp-Rb87}%
  \BibitemOpen
  \bibfield  {author} {\bibinfo {author} {\bibfnamefont {A.~J.}\ \bibnamefont {Olson}}, \bibinfo {author} {\bibfnamefont {S.-J.}\ \bibnamefont {Wang}}, \bibinfo {author} {\bibfnamefont {R.~J.}\ \bibnamefont {Niffenegger}}, \bibinfo {author} {\bibfnamefont {C.-H.}\ \bibnamefont {Li}}, \bibinfo {author} {\bibfnamefont {C.~H.}\ \bibnamefont {Greene}}, \ and\ \bibinfo {author} {\bibfnamefont {Y.~P.}\ \bibnamefont {Chen}},\ }\href {\doibase 10.1103/PhysRevA.90.013616} {\bibfield  {journal} {\bibinfo  {journal} {Phys. Rev. A}\ }\textbf {\bibinfo {volume} {90}},\ \bibinfo {pages} {013616} (\bibinfo {year} {2014})}\BibitemShut {NoStop}%
\bibitem [{\citenamefont {Luo}\ \emph {et~al.}(2016)\citenamefont {Luo}, \citenamefont {Wu}, \citenamefont {Chen}, \citenamefont {Guan}, \citenamefont {Gao}, \citenamefont {Xu}, \citenamefont {You},\ and\ \citenamefont {Wang}}]{Luo2016Jan-soc-exp-Rb87}%
  \BibitemOpen
  \bibfield  {author} {\bibinfo {author} {\bibfnamefont {X.}~\bibnamefont {Luo}}, \bibinfo {author} {\bibfnamefont {L.}~\bibnamefont {Wu}}, \bibinfo {author} {\bibfnamefont {J.}~\bibnamefont {Chen}}, \bibinfo {author} {\bibfnamefont {Q.}~\bibnamefont {Guan}}, \bibinfo {author} {\bibfnamefont {K.}~\bibnamefont {Gao}}, \bibinfo {author} {\bibfnamefont {Z.-F.}\ \bibnamefont {Xu}}, \bibinfo {author} {\bibfnamefont {L.}~\bibnamefont {You}}, \ and\ \bibinfo {author} {\bibfnamefont {R.}~\bibnamefont {Wang}},\ }\href {\doibase 10.1038/srep18983} {\bibfield  {journal} {\bibinfo  {journal} {Sci. Rep.}\ }\textbf {\bibinfo {volume} {6}},\ \bibinfo {pages} {18983} (\bibinfo {year} {2016})}\BibitemShut {NoStop}%
\bibitem [{\citenamefont {Wu}\ \emph {et~al.}(2016)\citenamefont {Wu}, \citenamefont {Zhang}, \citenamefont {Sun}, \citenamefont {Xu}, \citenamefont {Wang}, \citenamefont {Ji}, \citenamefont {Deng}, \citenamefont {Chen}, \citenamefont {Liu},\ and\ \citenamefont {Pan}}]{Wu2016Oct-soc-exp-Rb87}%
  \BibitemOpen
  \bibfield  {author} {\bibinfo {author} {\bibfnamefont {Z.}~\bibnamefont {Wu}}, \bibinfo {author} {\bibfnamefont {L.}~\bibnamefont {Zhang}}, \bibinfo {author} {\bibfnamefont {W.}~\bibnamefont {Sun}}, \bibinfo {author} {\bibfnamefont {X.-T.}\ \bibnamefont {Xu}}, \bibinfo {author} {\bibfnamefont {B.-Z.}\ \bibnamefont {Wang}}, \bibinfo {author} {\bibfnamefont {S.-C.}\ \bibnamefont {Ji}}, \bibinfo {author} {\bibfnamefont {Y.}~\bibnamefont {Deng}}, \bibinfo {author} {\bibfnamefont {S.}~\bibnamefont {Chen}}, \bibinfo {author} {\bibfnamefont {X.-J.}\ \bibnamefont {Liu}}, \ and\ \bibinfo {author} {\bibfnamefont {J.-W.}\ \bibnamefont {Pan}},\ }\href {\doibase 10.1126/science.aaf6689} {\bibfield  {journal} {\bibinfo  {journal} {Science}\ }\textbf {\bibinfo {volume} {354}},\ \bibinfo {pages} {83} (\bibinfo {year} {2016})}\BibitemShut {NoStop}%
\bibitem [{\citenamefont {Li}\ \emph {et~al.}(2016)\citenamefont {Li}, \citenamefont {Huang}, \citenamefont {Shteynas}, \citenamefont {Burchesky}, \citenamefont {Top}, \citenamefont {Su}, \citenamefont {Lee}, \citenamefont {Jamison},\ and\ \citenamefont {Ketterle}}]{Li2016Oct-soc-exp-Rb87}%
  \BibitemOpen
  \bibfield  {author} {\bibinfo {author} {\bibfnamefont {J.}~\bibnamefont {Li}}, \bibinfo {author} {\bibfnamefont {W.}~\bibnamefont {Huang}}, \bibinfo {author} {\bibfnamefont {B.}~\bibnamefont {Shteynas}}, \bibinfo {author} {\bibfnamefont {S.}~\bibnamefont {Burchesky}}, \bibinfo {author} {\bibfnamefont {F.~{\ifmmode\mbox{\c{C}}\else\c{C}\fi}.}\ \bibnamefont {Top}}, \bibinfo {author} {\bibfnamefont {E.}~\bibnamefont {Su}}, \bibinfo {author} {\bibfnamefont {J.}~\bibnamefont {Lee}}, \bibinfo {author} {\bibfnamefont {A.~O.}\ \bibnamefont {Jamison}}, \ and\ \bibinfo {author} {\bibfnamefont {W.}~\bibnamefont {Ketterle}},\ }\href {\doibase 10.1103/PhysRevLett.117.185301} {\bibfield  {journal} {\bibinfo  {journal} {Phys. Rev. Lett.}\ }\textbf {\bibinfo {volume} {117}},\ \bibinfo {pages} {185301} (\bibinfo {year} {2016})}\BibitemShut {NoStop}%
\bibitem [{\citenamefont {An}\ \emph {et~al.}(2017)\citenamefont {An}, \citenamefont {Meier},\ and\ \citenamefont {Gadway}}]{An2017Apr-soc-exp-Rb87}%
  \BibitemOpen
  \bibfield  {author} {\bibinfo {author} {\bibfnamefont {F.~A.}\ \bibnamefont {An}}, \bibinfo {author} {\bibfnamefont {E.~J.}\ \bibnamefont {Meier}}, \ and\ \bibinfo {author} {\bibfnamefont {B.}~\bibnamefont {Gadway}},\ }\href {\doibase 10.1126/sciadv.1602685} {\bibfield  {journal} {\bibinfo  {journal} {Sci. Adv.}\ }\textbf {\bibinfo {volume} {3}},\ \bibinfo {pages} {1602685} (\bibinfo {year} {2017})}\BibitemShut {NoStop}%
\bibitem [{\citenamefont {Li}\ \emph {et~al.}(2017)\citenamefont {Li}, \citenamefont {Lee}, \citenamefont {Huang}, \citenamefont {Burchesky}, \citenamefont {Shteynas}, \citenamefont {Top}, \citenamefont {Jamison},\ and\ \citenamefont {Ketterle}}]{Li2017Mar-soc-exp-Rb87}%
  \BibitemOpen
  \bibfield  {author} {\bibinfo {author} {\bibfnamefont {J.-R.}\ \bibnamefont {Li}}, \bibinfo {author} {\bibfnamefont {J.}~\bibnamefont {Lee}}, \bibinfo {author} {\bibfnamefont {W.}~\bibnamefont {Huang}}, \bibinfo {author} {\bibfnamefont {S.}~\bibnamefont {Burchesky}}, \bibinfo {author} {\bibfnamefont {B.}~\bibnamefont {Shteynas}}, \bibinfo {author} {\bibfnamefont {F.~{\ifmmode\mbox{\c{C}}\else\c{C}\fi}.}\ \bibnamefont {Top}}, \bibinfo {author} {\bibfnamefont {A.~O.}\ \bibnamefont {Jamison}}, \ and\ \bibinfo {author} {\bibfnamefont {W.}~\bibnamefont {Ketterle}},\ }\href {\doibase 10.1038/nature21431} {\bibfield  {journal} {\bibinfo  {journal} {Nature}\ }\textbf {\bibinfo {volume} {543}},\ \bibinfo {pages} {91} (\bibinfo {year} {2017})}\BibitemShut {NoStop}%
\bibitem [{\citenamefont {Sun}\ \emph {et~al.}(2018)\citenamefont {Sun}, \citenamefont {Wang}, \citenamefont {Xu}, \citenamefont {Yi}, \citenamefont {Zhang}, \citenamefont {Wu}, \citenamefont {Deng}, \citenamefont {Liu}, \citenamefont {Chen},\ and\ \citenamefont {Pan}}]{Sun2018Oct-soc-exp-Rb87}%
  \BibitemOpen
  \bibfield  {author} {\bibinfo {author} {\bibfnamefont {W.}~\bibnamefont {Sun}}, \bibinfo {author} {\bibfnamefont {B.-Z.}\ \bibnamefont {Wang}}, \bibinfo {author} {\bibfnamefont {X.-T.}\ \bibnamefont {Xu}}, \bibinfo {author} {\bibfnamefont {C.-R.}\ \bibnamefont {Yi}}, \bibinfo {author} {\bibfnamefont {L.}~\bibnamefont {Zhang}}, \bibinfo {author} {\bibfnamefont {Z.}~\bibnamefont {Wu}}, \bibinfo {author} {\bibfnamefont {Y.}~\bibnamefont {Deng}}, \bibinfo {author} {\bibfnamefont {X.-J.}\ \bibnamefont {Liu}}, \bibinfo {author} {\bibfnamefont {S.}~\bibnamefont {Chen}}, \ and\ \bibinfo {author} {\bibfnamefont {J.-W.}\ \bibnamefont {Pan}},\ }\href {\doibase 10.1103/PhysRevLett.121.150401} {\bibfield  {journal} {\bibinfo  {journal} {Phys. Rev. Lett.}\ }\textbf {\bibinfo {volume} {121}},\ \bibinfo {pages} {150401} (\bibinfo {year} {2018})}\BibitemShut {NoStop}%
\bibitem [{\citenamefont {Wang}\ \emph {et~al.}(2021)\citenamefont {Wang}, \citenamefont {Cheng}, \citenamefont {Wang}, \citenamefont {Zhang}, \citenamefont {Lu}, \citenamefont {Yi}, \citenamefont {Niu}, \citenamefont {Deng}, \citenamefont {Liu}, \citenamefont {Chen},\ and\ \citenamefont {Pan}}]{Wang2021Apr-soc-exp-Rb87}%
  \BibitemOpen
  \bibfield  {author} {\bibinfo {author} {\bibfnamefont {Z.-Y.}\ \bibnamefont {Wang}}, \bibinfo {author} {\bibfnamefont {X.-C.}\ \bibnamefont {Cheng}}, \bibinfo {author} {\bibfnamefont {B.-Z.}\ \bibnamefont {Wang}}, \bibinfo {author} {\bibfnamefont {J.-Y.}\ \bibnamefont {Zhang}}, \bibinfo {author} {\bibfnamefont {Y.-H.}\ \bibnamefont {Lu}}, \bibinfo {author} {\bibfnamefont {C.-R.}\ \bibnamefont {Yi}}, \bibinfo {author} {\bibfnamefont {S.}~\bibnamefont {Niu}}, \bibinfo {author} {\bibfnamefont {Y.}~\bibnamefont {Deng}}, \bibinfo {author} {\bibfnamefont {X.-J.}\ \bibnamefont {Liu}}, \bibinfo {author} {\bibfnamefont {S.}~\bibnamefont {Chen}}, \ and\ \bibinfo {author} {\bibfnamefont {J.-W.}\ \bibnamefont {Pan}},\ }\href {\doibase 10.1126/science.abc0105} {\bibfield  {journal} {\bibinfo  {journal} {Science}\ }\textbf {\bibinfo {volume} {372}},\ \bibinfo {pages} {271} (\bibinfo {year} {2021})}\BibitemShut {NoStop}%
\bibitem [{\citenamefont {Huang}\ \emph {et~al.}(2016)\citenamefont {Huang}, \citenamefont {Meng}, \citenamefont {Wang}, \citenamefont {Peng}, \citenamefont {Zhang}, \citenamefont {Chen}, \citenamefont {Li}, \citenamefont {Zhou},\ and\ \citenamefont {Zhang}}]{Huang2016Jun-nat-phys-soc-exp-K40}%
  \BibitemOpen
  \bibfield  {author} {\bibinfo {author} {\bibfnamefont {L.}~\bibnamefont {Huang}}, \bibinfo {author} {\bibfnamefont {Z.}~\bibnamefont {Meng}}, \bibinfo {author} {\bibfnamefont {P.}~\bibnamefont {Wang}}, \bibinfo {author} {\bibfnamefont {P.}~\bibnamefont {Peng}}, \bibinfo {author} {\bibfnamefont {S.-L.}\ \bibnamefont {Zhang}}, \bibinfo {author} {\bibfnamefont {L.}~\bibnamefont {Chen}}, \bibinfo {author} {\bibfnamefont {D.}~\bibnamefont {Li}}, \bibinfo {author} {\bibfnamefont {Q.}~\bibnamefont {Zhou}}, \ and\ \bibinfo {author} {\bibfnamefont {J.}~\bibnamefont {Zhang}},\ }\href {\doibase 10.1038/nphys3672} {\bibfield  {journal} {\bibinfo  {journal} {Nat. Phys.}\ }\textbf {\bibinfo {volume} {12}},\ \bibinfo {pages} {540} (\bibinfo {year} {2016})}\BibitemShut {NoStop}%
\bibitem [{\citenamefont {Meng}\ \emph {et~al.}(2016)\citenamefont {Meng}, \citenamefont {Huang}, \citenamefont {Peng}, \citenamefont {Li}, \citenamefont {Chen}, \citenamefont {Xu}, \citenamefont {Zhang}, \citenamefont {Wang},\ and\ \citenamefont {Zhang}}]{Meng2016Dec-soc-exp-K40}%
  \BibitemOpen
  \bibfield  {author} {\bibinfo {author} {\bibfnamefont {Z.}~\bibnamefont {Meng}}, \bibinfo {author} {\bibfnamefont {L.}~\bibnamefont {Huang}}, \bibinfo {author} {\bibfnamefont {P.}~\bibnamefont {Peng}}, \bibinfo {author} {\bibfnamefont {D.}~\bibnamefont {Li}}, \bibinfo {author} {\bibfnamefont {L.}~\bibnamefont {Chen}}, \bibinfo {author} {\bibfnamefont {Y.}~\bibnamefont {Xu}}, \bibinfo {author} {\bibfnamefont {C.}~\bibnamefont {Zhang}}, \bibinfo {author} {\bibfnamefont {P.}~\bibnamefont {Wang}}, \ and\ \bibinfo {author} {\bibfnamefont {J.}~\bibnamefont {Zhang}},\ }\href {\doibase 10.1103/PhysRevLett.117.235304} {\bibfield  {journal} {\bibinfo  {journal} {Phys. Rev. Lett.}\ }\textbf {\bibinfo {volume} {117}},\ \bibinfo {pages} {235304} (\bibinfo {year} {2016})}\BibitemShut {NoStop}%
\bibitem [{\citenamefont {Cheuk}\ \emph {et~al.}(2012)\citenamefont {Cheuk}, \citenamefont {Sommer}, \citenamefont {Hadzibabic}, \citenamefont {Yefsah}, \citenamefont {Bakr},\ and\ \citenamefont {Zwierlein}}]{Cheuk2012Aug-soc-exp-Li6}%
  \BibitemOpen
  \bibfield  {author} {\bibinfo {author} {\bibfnamefont {L.~W.}\ \bibnamefont {Cheuk}}, \bibinfo {author} {\bibfnamefont {A.~T.}\ \bibnamefont {Sommer}}, \bibinfo {author} {\bibfnamefont {Z.}~\bibnamefont {Hadzibabic}}, \bibinfo {author} {\bibfnamefont {T.}~\bibnamefont {Yefsah}}, \bibinfo {author} {\bibfnamefont {W.~S.}\ \bibnamefont {Bakr}}, \ and\ \bibinfo {author} {\bibfnamefont {M.~W.}\ \bibnamefont {Zwierlein}},\ }\href {\doibase 10.1103/PhysRevLett.109.095302} {\bibfield  {journal} {\bibinfo  {journal} {Phys. Rev. Lett.}\ }\textbf {\bibinfo {volume} {109}},\ \bibinfo {pages} {095302} (\bibinfo {year} {2012})}\BibitemShut {NoStop}%
\bibitem [{\citenamefont {Kolkowitz}\ \emph {et~al.}(2017)\citenamefont {Kolkowitz}, \citenamefont {Bromley}, \citenamefont {Bothwell}, \citenamefont {Wall}, \citenamefont {Marti}, \citenamefont {Koller}, \citenamefont {Zhang}, \citenamefont {Rey},\ and\ \citenamefont {Ye}}]{Kolkowitz2017Feb-soc-exp-Sr87}%
  \BibitemOpen
  \bibfield  {author} {\bibinfo {author} {\bibfnamefont {S.}~\bibnamefont {Kolkowitz}}, \bibinfo {author} {\bibfnamefont {S.~L.}\ \bibnamefont {Bromley}}, \bibinfo {author} {\bibfnamefont {T.}~\bibnamefont {Bothwell}}, \bibinfo {author} {\bibfnamefont {M.~L.}\ \bibnamefont {Wall}}, \bibinfo {author} {\bibfnamefont {G.~E.}\ \bibnamefont {Marti}}, \bibinfo {author} {\bibfnamefont {A.~P.}\ \bibnamefont {Koller}}, \bibinfo {author} {\bibfnamefont {X.}~\bibnamefont {Zhang}}, \bibinfo {author} {\bibfnamefont {A.~M.}\ \bibnamefont {Rey}}, \ and\ \bibinfo {author} {\bibfnamefont {J.}~\bibnamefont {Ye}},\ }\href {\doibase 10.1038/nature20811} {\bibfield  {journal} {\bibinfo  {journal} {Nature}\ }\textbf {\bibinfo {volume} {542}},\ \bibinfo {pages} {66} (\bibinfo {year} {2017})}\BibitemShut {NoStop}%
\bibitem [{\citenamefont {Aeppli}\ \emph {et~al.}(2022)\citenamefont {Aeppli}, \citenamefont {Chu}, \citenamefont {Bothwell}, \citenamefont {Kennedy}, \citenamefont {Kedar}, \citenamefont {He}, \citenamefont {Rey},\ and\ \citenamefont {Ye}}]{Aeppli2022Oct-soc-exp-Sr87}%
  \BibitemOpen
  \bibfield  {author} {\bibinfo {author} {\bibfnamefont {A.}~\bibnamefont {Aeppli}}, \bibinfo {author} {\bibfnamefont {A.}~\bibnamefont {Chu}}, \bibinfo {author} {\bibfnamefont {T.}~\bibnamefont {Bothwell}}, \bibinfo {author} {\bibfnamefont {C.~J.}\ \bibnamefont {Kennedy}}, \bibinfo {author} {\bibfnamefont {D.}~\bibnamefont {Kedar}}, \bibinfo {author} {\bibfnamefont {P.}~\bibnamefont {He}}, \bibinfo {author} {\bibfnamefont {A.~M.}\ \bibnamefont {Rey}}, \ and\ \bibinfo {author} {\bibfnamefont {J.}~\bibnamefont {Ye}},\ }\href {\doibase 10.1126/sciadv.adc9242} {\bibfield  {journal} {\bibinfo  {journal} {Sci. Adv.}\ }\textbf {\bibinfo {volume} {8}},\ \bibinfo {pages} {adc924} (\bibinfo {year} {2022})}\BibitemShut {NoStop}%
\bibitem [{\citenamefont {Liang}\ \emph {et~al.}(2023)\citenamefont {Liang}, \citenamefont {Wei}, \citenamefont {Zhang}, \citenamefont {Wang}, \citenamefont {Zhang}, \citenamefont {Wang}, \citenamefont {Qi}, \citenamefont {Liu},\ and\ \citenamefont {Zhang}}]{Liang2023Jan-soc-exp-Sr87}%
  \BibitemOpen
  \bibfield  {author} {\bibinfo {author} {\bibfnamefont {M.-C.}\ \bibnamefont {Liang}}, \bibinfo {author} {\bibfnamefont {Y.-D.}\ \bibnamefont {Wei}}, \bibinfo {author} {\bibfnamefont {L.}~\bibnamefont {Zhang}}, \bibinfo {author} {\bibfnamefont {X.-J.}\ \bibnamefont {Wang}}, \bibinfo {author} {\bibfnamefont {H.}~\bibnamefont {Zhang}}, \bibinfo {author} {\bibfnamefont {W.-W.}\ \bibnamefont {Wang}}, \bibinfo {author} {\bibfnamefont {W.}~\bibnamefont {Qi}}, \bibinfo {author} {\bibfnamefont {X.-J.}\ \bibnamefont {Liu}}, \ and\ \bibinfo {author} {\bibfnamefont {X.}~\bibnamefont {Zhang}},\ }\href {\doibase 10.1103/PhysRevResearch.5.L012006} {\bibfield  {journal} {\bibinfo  {journal} {Phys. Rev. Res.}\ }\textbf {\bibinfo {volume} {5}},\ \bibinfo {pages} {L012006} (\bibinfo {year} {2023})}\BibitemShut {NoStop}%
\bibitem [{\citenamefont {Burdick}\ \emph {et~al.}(2016)\citenamefont {Burdick}, \citenamefont {Tang},\ and\ \citenamefont {Lev}}]{Burdick2016Aug-soc-Dy161}%
  \BibitemOpen
  \bibfield  {author} {\bibinfo {author} {\bibfnamefont {N.~Q.}\ \bibnamefont {Burdick}}, \bibinfo {author} {\bibfnamefont {Y.}~\bibnamefont {Tang}}, \ and\ \bibinfo {author} {\bibfnamefont {B.~L.}\ \bibnamefont {Lev}},\ }\href {\doibase 10.1103/PhysRevX.6.031022} {\bibfield  {journal} {\bibinfo  {journal} {Phys. Rev. X}\ }\textbf {\bibinfo {volume} {6}},\ \bibinfo {pages} {031022} (\bibinfo {year} {2016})}\BibitemShut {NoStop}%
\bibitem [{\citenamefont {Livi}\ \emph {et~al.}(2016)\citenamefont {Livi}, \citenamefont {Cappellini}, \citenamefont {Diem}, \citenamefont {Franchi}, \citenamefont {Clivati}, \citenamefont {Frittelli}, \citenamefont {Levi}, \citenamefont {Calonico}, \citenamefont {Catani}, \citenamefont {Inguscio},\ and\ \citenamefont {Fallani}}]{Livi2016Nov-soc-exp-Yb173}%
  \BibitemOpen
  \bibfield  {author} {\bibinfo {author} {\bibfnamefont {L.~F.}\ \bibnamefont {Livi}}, \bibinfo {author} {\bibfnamefont {G.}~\bibnamefont {Cappellini}}, \bibinfo {author} {\bibfnamefont {M.}~\bibnamefont {Diem}}, \bibinfo {author} {\bibfnamefont {L.}~\bibnamefont {Franchi}}, \bibinfo {author} {\bibfnamefont {C.}~\bibnamefont {Clivati}}, \bibinfo {author} {\bibfnamefont {M.}~\bibnamefont {Frittelli}}, \bibinfo {author} {\bibfnamefont {F.}~\bibnamefont {Levi}}, \bibinfo {author} {\bibfnamefont {D.}~\bibnamefont {Calonico}}, \bibinfo {author} {\bibfnamefont {J.}~\bibnamefont {Catani}}, \bibinfo {author} {\bibfnamefont {M.}~\bibnamefont {Inguscio}}, \ and\ \bibinfo {author} {\bibfnamefont {L.}~\bibnamefont {Fallani}},\ }\href {\doibase 10.1103/PhysRevLett.117.220401} {\bibfield  {journal} {\bibinfo  {journal} {Phys. Rev. Lett.}\ }\textbf {\bibinfo {volume} {117}},\ \bibinfo {pages} {220401} (\bibinfo {year} {2016})}\BibitemShut {NoStop}%
\bibitem [{\citenamefont {Song}\ \emph {et~al.}(2016)\citenamefont {Song}, \citenamefont {He}, \citenamefont {Zhang}, \citenamefont {Hajiyev}, \citenamefont {Huang}, \citenamefont {Liu},\ and\ \citenamefont {Jo}}]{Song2016Dec-soc-exp-Yb173}%
  \BibitemOpen
  \bibfield  {author} {\bibinfo {author} {\bibfnamefont {B.}~\bibnamefont {Song}}, \bibinfo {author} {\bibfnamefont {C.}~\bibnamefont {He}}, \bibinfo {author} {\bibfnamefont {S.}~\bibnamefont {Zhang}}, \bibinfo {author} {\bibfnamefont {E.}~\bibnamefont {Hajiyev}}, \bibinfo {author} {\bibfnamefont {W.}~\bibnamefont {Huang}}, \bibinfo {author} {\bibfnamefont {X.-J.}\ \bibnamefont {Liu}}, \ and\ \bibinfo {author} {\bibfnamefont {G.-B.}\ \bibnamefont {Jo}},\ }\href {\doibase 10.1103/PhysRevA.94.061604} {\bibfield  {journal} {\bibinfo  {journal} {Phys. Rev. A}\ }\textbf {\bibinfo {volume} {94}},\ \bibinfo {pages} {061604} (\bibinfo {year} {2016})}\BibitemShut {NoStop}%
\bibitem [{\citenamefont {Song}\ \emph {et~al.}(2019)\citenamefont {Song}, \citenamefont {He}, \citenamefont {Niu}, \citenamefont {Zhang}, \citenamefont {Ren}, \citenamefont {Liu},\ and\ \citenamefont {Jo}}]{Song2019Sep-soc-exp-Yb173}%
  \BibitemOpen
  \bibfield  {author} {\bibinfo {author} {\bibfnamefont {B.}~\bibnamefont {Song}}, \bibinfo {author} {\bibfnamefont {C.}~\bibnamefont {He}}, \bibinfo {author} {\bibfnamefont {S.}~\bibnamefont {Niu}}, \bibinfo {author} {\bibfnamefont {L.}~\bibnamefont {Zhang}}, \bibinfo {author} {\bibfnamefont {Z.}~\bibnamefont {Ren}}, \bibinfo {author} {\bibfnamefont {X.-J.}\ \bibnamefont {Liu}}, \ and\ \bibinfo {author} {\bibfnamefont {G.-B.}\ \bibnamefont {Jo}},\ }\href {\doibase 10.1038/s41567-019-0564-y} {\bibfield  {journal} {\bibinfo  {journal} {Nat. Phys.}\ }\textbf {\bibinfo {volume} {15}},\ \bibinfo {pages} {911} (\bibinfo {year} {2019})}\BibitemShut {NoStop}%
\bibitem [{\citenamefont {Liu}\ \emph {et~al.}(2014)\citenamefont {Liu}, \citenamefont {Law},\ and\ \citenamefont {Ng}}]{Liu2014Feb-soc-theo}%
  \BibitemOpen
  \bibfield  {author} {\bibinfo {author} {\bibfnamefont {X.-J.}\ \bibnamefont {Liu}}, \bibinfo {author} {\bibfnamefont {K.~T.}\ \bibnamefont {Law}}, \ and\ \bibinfo {author} {\bibfnamefont {T.~K.}\ \bibnamefont {Ng}},\ }\href {\doibase 10.1103/PhysRevLett.112.086401} {\bibfield  {journal} {\bibinfo  {journal} {Phys. Rev. Lett.}\ }\textbf {\bibinfo {volume} {112}},\ \bibinfo {pages} {086401} (\bibinfo {year} {2014})}\BibitemShut {NoStop}%
\bibitem [{\citenamefont {Struck}\ \emph {et~al.}(2014)\citenamefont {Struck}, \citenamefont {Simonet},\ and\ \citenamefont {Sengstock}}]{Struck2014Sep-soc-theo}%
  \BibitemOpen
  \bibfield  {author} {\bibinfo {author} {\bibfnamefont {J.}~\bibnamefont {Struck}}, \bibinfo {author} {\bibfnamefont {J.}~\bibnamefont {Simonet}}, \ and\ \bibinfo {author} {\bibfnamefont {K.}~\bibnamefont {Sengstock}},\ }\href {\doibase 10.1103/PhysRevA.90.031601} {\bibfield  {journal} {\bibinfo  {journal} {Phys. Rev. A}\ }\textbf {\bibinfo {volume} {90}},\ \bibinfo {pages} {031601} (\bibinfo {year} {2014})}\BibitemShut {NoStop}%
\bibitem [{\citenamefont {Wall}\ \emph {et~al.}(2016)\citenamefont {Wall}, \citenamefont {Koller}, \citenamefont {Li}, \citenamefont {Zhang}, \citenamefont {Cooper}, \citenamefont {Ye},\ and\ \citenamefont {Rey}}]{Wall2016Jan-soc-theo}%
  \BibitemOpen
  \bibfield  {author} {\bibinfo {author} {\bibfnamefont {M.~L.}\ \bibnamefont {Wall}}, \bibinfo {author} {\bibfnamefont {A.~P.}\ \bibnamefont {Koller}}, \bibinfo {author} {\bibfnamefont {S.}~\bibnamefont {Li}}, \bibinfo {author} {\bibfnamefont {X.}~\bibnamefont {Zhang}}, \bibinfo {author} {\bibfnamefont {N.~R.}\ \bibnamefont {Cooper}}, \bibinfo {author} {\bibfnamefont {J.}~\bibnamefont {Ye}}, \ and\ \bibinfo {author} {\bibfnamefont {A.~M.}\ \bibnamefont {Rey}},\ }\href {\doibase 10.1103/PhysRevLett.116.035301} {\bibfield  {journal} {\bibinfo  {journal} {Phys. Rev. Lett.}\ }\textbf {\bibinfo {volume} {116}},\ \bibinfo {pages} {035301} (\bibinfo {year} {2016})}\BibitemShut {NoStop}%
\bibitem [{\citenamefont {Zhang}\ \emph {et~al.}(2008)\citenamefont {Zhang}, \citenamefont {Tewari}, \citenamefont {Lutchyn},\ and\ \citenamefont {Das~Sarma}}]{Zhang2008Oct-soc-appli}%
  \BibitemOpen
  \bibfield  {author} {\bibinfo {author} {\bibfnamefont {C.}~\bibnamefont {Zhang}}, \bibinfo {author} {\bibfnamefont {S.}~\bibnamefont {Tewari}}, \bibinfo {author} {\bibfnamefont {R.~M.}\ \bibnamefont {Lutchyn}}, \ and\ \bibinfo {author} {\bibfnamefont {S.}~\bibnamefont {Das~Sarma}},\ }\href {\doibase 10.1103/PhysRevLett.101.160401} {\bibfield  {journal} {\bibinfo  {journal} {Phys. Rev. Lett.}\ }\textbf {\bibinfo {volume} {101}},\ \bibinfo {pages} {160401} (\bibinfo {year} {2008})}\BibitemShut {NoStop}%
\bibitem [{\citenamefont {Liu}\ \emph {et~al.}(2009)\citenamefont {Liu}, \citenamefont {Borunda}, \citenamefont {Liu},\ and\ \citenamefont {Sinova}}]{Liu2009Jan-soc-appli}%
  \BibitemOpen
  \bibfield  {author} {\bibinfo {author} {\bibfnamefont {X.-J.}\ \bibnamefont {Liu}}, \bibinfo {author} {\bibfnamefont {M.~F.}\ \bibnamefont {Borunda}}, \bibinfo {author} {\bibfnamefont {X.}~\bibnamefont {Liu}}, \ and\ \bibinfo {author} {\bibfnamefont {J.}~\bibnamefont {Sinova}},\ }\href {\doibase 10.1103/PhysRevLett.102.046402} {\bibfield  {journal} {\bibinfo  {journal} {Phys. Rev. Lett.}\ }\textbf {\bibinfo {volume} {102}},\ \bibinfo {pages} {046402} (\bibinfo {year} {2009})}\BibitemShut {NoStop}%
\bibitem [{\citenamefont {Wang}\ \emph {et~al.}(2010)\citenamefont {Wang}, \citenamefont {Gao}, \citenamefont {Jian},\ and\ \citenamefont {Zhai}}]{Wang2010Oct-soc-appli}%
  \BibitemOpen
  \bibfield  {author} {\bibinfo {author} {\bibfnamefont {C.}~\bibnamefont {Wang}}, \bibinfo {author} {\bibfnamefont {C.}~\bibnamefont {Gao}}, \bibinfo {author} {\bibfnamefont {C.-M.}\ \bibnamefont {Jian}}, \ and\ \bibinfo {author} {\bibfnamefont {H.}~\bibnamefont {Zhai}},\ }\href {\doibase 10.1103/PhysRevLett.105.160403} {\bibfield  {journal} {\bibinfo  {journal} {Phys. Rev. Lett.}\ }\textbf {\bibinfo {volume} {105}},\ \bibinfo {pages} {160403} (\bibinfo {year} {2010})}\BibitemShut {NoStop}%
\bibitem [{\citenamefont {Gong}\ \emph {et~al.}(2011)\citenamefont {Gong}, \citenamefont {Tewari},\ and\ \citenamefont {Zhang}}]{Gong2011Nov-soc-appli}%
  \BibitemOpen
  \bibfield  {author} {\bibinfo {author} {\bibfnamefont {M.}~\bibnamefont {Gong}}, \bibinfo {author} {\bibfnamefont {S.}~\bibnamefont {Tewari}}, \ and\ \bibinfo {author} {\bibfnamefont {C.}~\bibnamefont {Zhang}},\ }\href {\doibase 10.1103/PhysRevLett.107.195303} {\bibfield  {journal} {\bibinfo  {journal} {Phys. Rev. Lett.}\ }\textbf {\bibinfo {volume} {107}},\ \bibinfo {pages} {195303} (\bibinfo {year} {2011})}\BibitemShut {NoStop}%
\bibitem [{\citenamefont {Zhang}\ \emph {et~al.}(2012)\citenamefont {Zhang}, \citenamefont {Xue}, \citenamefont {Yan}, \citenamefont {Wang},\ and\ \citenamefont {Zhu}}]{DWZhang2012-soc-appli}%
  \BibitemOpen
  \bibfield  {author} {\bibinfo {author} {\bibfnamefont {D.-W.}\ \bibnamefont {Zhang}}, \bibinfo {author} {\bibfnamefont {Z.-Y.}\ \bibnamefont {Xue}}, \bibinfo {author} {\bibfnamefont {H.}~\bibnamefont {Yan}}, \bibinfo {author} {\bibfnamefont {Z.~D.}\ \bibnamefont {Wang}}, \ and\ \bibinfo {author} {\bibfnamefont {S.-L.}\ \bibnamefont {Zhu}},\ }\href {\doibase 10.1103/physreva.85.013628} {\bibfield  {journal} {\bibinfo  {journal} {Phys. Rev. A}\ }\textbf {\bibinfo {volume} {85}},\ \bibinfo {pages} {013628} (\bibinfo {year} {2012})}\BibitemShut {NoStop}%
\bibitem [{\citenamefont {Ozawa}\ and\ \citenamefont {Baym}(2013)}]{Ozawa2013Feb-soc-appli}%
  \BibitemOpen
  \bibfield  {author} {\bibinfo {author} {\bibfnamefont {T.}~\bibnamefont {Ozawa}}\ and\ \bibinfo {author} {\bibfnamefont {G.}~\bibnamefont {Baym}},\ }\href {\doibase 10.1103/PhysRevLett.110.085304} {\bibfield  {journal} {\bibinfo  {journal} {Phys. Rev. Lett.}\ }\textbf {\bibinfo {volume} {110}},\ \bibinfo {pages} {085304} (\bibinfo {year} {2013})}\BibitemShut {NoStop}%
\bibitem [{\citenamefont {Zheng}\ \emph {et~al.}(2013)\citenamefont {Zheng}, \citenamefont {Gong}, \citenamefont {Zou}, \citenamefont {Zhang},\ and\ \citenamefont {Guo}}]{Zheng2013Mar-soc-appli}%
  \BibitemOpen
  \bibfield  {author} {\bibinfo {author} {\bibfnamefont {Z.}~\bibnamefont {Zheng}}, \bibinfo {author} {\bibfnamefont {M.}~\bibnamefont {Gong}}, \bibinfo {author} {\bibfnamefont {X.}~\bibnamefont {Zou}}, \bibinfo {author} {\bibfnamefont {C.}~\bibnamefont {Zhang}}, \ and\ \bibinfo {author} {\bibfnamefont {G.}~\bibnamefont {Guo}},\ }\href {\doibase 10.1103/PhysRevA.87.031602} {\bibfield  {journal} {\bibinfo  {journal} {Phys. Rev. A}\ }\textbf {\bibinfo {volume} {87}},\ \bibinfo {pages} {031602} (\bibinfo {year} {2013})}\BibitemShut {NoStop}%
\bibitem [{\citenamefont {Deng}\ \emph {et~al.}(2014{\natexlab{b}})\citenamefont {Deng}, \citenamefont {Cheng}, \citenamefont {Jing},\ and\ \citenamefont {Yi}}]{Deng2014Apr-soc-appli}%
  \BibitemOpen
  \bibfield  {author} {\bibinfo {author} {\bibfnamefont {Y.}~\bibnamefont {Deng}}, \bibinfo {author} {\bibfnamefont {J.}~\bibnamefont {Cheng}}, \bibinfo {author} {\bibfnamefont {H.}~\bibnamefont {Jing}}, \ and\ \bibinfo {author} {\bibfnamefont {S.}~\bibnamefont {Yi}},\ }\href {\doibase 10.1103/PhysRevLett.112.143007} {\bibfield  {journal} {\bibinfo  {journal} {Phys. Rev. Lett.}\ }\textbf {\bibinfo {volume} {112}},\ \bibinfo {pages} {143007} (\bibinfo {year} {2014}{\natexlab{b}})}\BibitemShut {NoStop}%
\bibitem [{\citenamefont {Xu}\ and\ \citenamefont {Zhang}(2015)}]{Xu2015Mar-soc-appli}%
  \BibitemOpen
  \bibfield  {author} {\bibinfo {author} {\bibfnamefont {Y.}~\bibnamefont {Xu}}\ and\ \bibinfo {author} {\bibfnamefont {C.}~\bibnamefont {Zhang}},\ }\href {\doibase 10.1103/PhysRevLett.114.110401} {\bibfield  {journal} {\bibinfo  {journal} {Phys. Rev. Lett.}\ }\textbf {\bibinfo {volume} {114}},\ \bibinfo {pages} {110401} (\bibinfo {year} {2015})}\BibitemShut {NoStop}%
\bibitem [{\citenamefont {Sun}\ \emph {et~al.}(2015)\citenamefont {Sun}, \citenamefont {Ye},\ and\ \citenamefont {Liu}}]{Sun2015Oct-soc-appli}%
  \BibitemOpen
  \bibfield  {author} {\bibinfo {author} {\bibfnamefont {F.}~\bibnamefont {Sun}}, \bibinfo {author} {\bibfnamefont {J.}~\bibnamefont {Ye}}, \ and\ \bibinfo {author} {\bibfnamefont {W.-M.}\ \bibnamefont {Liu}},\ }\href {\doibase 10.1103/PhysRevA.92.043609} {\bibfield  {journal} {\bibinfo  {journal} {Phys. Rev. A}\ }\textbf {\bibinfo {volume} {92}},\ \bibinfo {pages} {043609} (\bibinfo {year} {2015})}\BibitemShut {NoStop}%
\bibitem [{\citenamefont {Deng}\ \emph {et~al.}(2017)\citenamefont {Deng}, \citenamefont {Shi}, \citenamefont {Hu}, \citenamefont {You},\ and\ \citenamefont {Yi}}]{Deng2017Feb-soc-appli}%
  \BibitemOpen
  \bibfield  {author} {\bibinfo {author} {\bibfnamefont {Y.}~\bibnamefont {Deng}}, \bibinfo {author} {\bibfnamefont {T.}~\bibnamefont {Shi}}, \bibinfo {author} {\bibfnamefont {H.}~\bibnamefont {Hu}}, \bibinfo {author} {\bibfnamefont {L.}~\bibnamefont {You}}, \ and\ \bibinfo {author} {\bibfnamefont {S.}~\bibnamefont {Yi}},\ }\href {\doibase 10.1103/PhysRevA.95.023611} {\bibfield  {journal} {\bibinfo  {journal} {Phys. Rev. A}\ }\textbf {\bibinfo {volume} {95}},\ \bibinfo {pages} {023611} (\bibinfo {year} {2017})}\BibitemShut {NoStop}%
\bibitem [{\citenamefont {Zhou}\ \emph {et~al.}(2017)\citenamefont {Zhou}, \citenamefont {Pan}, \citenamefont {Yi}, \citenamefont {Chen},\ and\ \citenamefont {Jia}}]{Zhou2017Aug-soc-appli}%
  \BibitemOpen
  \bibfield  {author} {\bibinfo {author} {\bibfnamefont {X.}~\bibnamefont {Zhou}}, \bibinfo {author} {\bibfnamefont {J.-S.}\ \bibnamefont {Pan}}, \bibinfo {author} {\bibfnamefont {W.}~\bibnamefont {Yi}}, \bibinfo {author} {\bibfnamefont {G.}~\bibnamefont {Chen}}, \ and\ \bibinfo {author} {\bibfnamefont {S.}~\bibnamefont {Jia}},\ }\href {\doibase 10.1103/PhysRevA.96.023627} {\bibfield  {journal} {\bibinfo  {journal} {Phys. Rev. A}\ }\textbf {\bibinfo {volume} {96}},\ \bibinfo {pages} {023627} (\bibinfo {year} {2017})}\BibitemShut {NoStop}%
\bibitem [{\citenamefont {Chen}\ \emph {et~al.}(2018)\citenamefont {Chen}, \citenamefont {Wang}, \citenamefont {Li}, \citenamefont {Liu},\ and\ \citenamefont {Hu}}]{Chen2018Jul-soc-appli}%
  \BibitemOpen
  \bibfield  {author} {\bibinfo {author} {\bibfnamefont {X.-L.}\ \bibnamefont {Chen}}, \bibinfo {author} {\bibfnamefont {J.}~\bibnamefont {Wang}}, \bibinfo {author} {\bibfnamefont {Y.}~\bibnamefont {Li}}, \bibinfo {author} {\bibfnamefont {X.-J.}\ \bibnamefont {Liu}}, \ and\ \bibinfo {author} {\bibfnamefont {H.}~\bibnamefont {Hu}},\ }\href {\doibase 10.1103/PhysRevA.98.013614} {\bibfield  {journal} {\bibinfo  {journal} {Phys. Rev. A}\ }\textbf {\bibinfo {volume} {98}},\ \bibinfo {pages} {013614} (\bibinfo {year} {2018})}\BibitemShut {NoStop}%
\bibitem [{\citenamefont {Tang}\ and\ \citenamefont {Zhang}(2018)}]{Tang2018Sep-soc-appli}%
  \BibitemOpen
  \bibfield  {author} {\bibinfo {author} {\bibfnamefont {W.~H.}\ \bibnamefont {Tang}}\ and\ \bibinfo {author} {\bibfnamefont {S.}~\bibnamefont {Zhang}},\ }\href {\doibase 10.1103/PhysRevLett.121.120403} {\bibfield  {journal} {\bibinfo  {journal} {Phys. Rev. Lett.}\ }\textbf {\bibinfo {volume} {121}},\ \bibinfo {pages} {120403} (\bibinfo {year} {2018})}\BibitemShut {NoStop}%
\bibitem [{\citenamefont {Li}\ \emph {et~al.}(2019)\citenamefont {Li}, \citenamefont {Qu}, \citenamefont {Niffenegger}, \citenamefont {Wang}, \citenamefont {He}, \citenamefont {Blasing}, \citenamefont {Olson}, \citenamefont {Greene}, \citenamefont {Lyanda-Geller}, \citenamefont {Zhou}, \citenamefont {Zhang},\ and\ \citenamefont {Chen}}]{Li2019Jan-soc-appli}%
  \BibitemOpen
  \bibfield  {author} {\bibinfo {author} {\bibfnamefont {C.-H.}\ \bibnamefont {Li}}, \bibinfo {author} {\bibfnamefont {C.}~\bibnamefont {Qu}}, \bibinfo {author} {\bibfnamefont {R.~J.}\ \bibnamefont {Niffenegger}}, \bibinfo {author} {\bibfnamefont {S.-J.}\ \bibnamefont {Wang}}, \bibinfo {author} {\bibfnamefont {M.}~\bibnamefont {He}}, \bibinfo {author} {\bibfnamefont {D.~B.}\ \bibnamefont {Blasing}}, \bibinfo {author} {\bibfnamefont {A.~J.}\ \bibnamefont {Olson}}, \bibinfo {author} {\bibfnamefont {C.~H.}\ \bibnamefont {Greene}}, \bibinfo {author} {\bibfnamefont {Y.}~\bibnamefont {Lyanda-Geller}}, \bibinfo {author} {\bibfnamefont {Q.}~\bibnamefont {Zhou}}, \bibinfo {author} {\bibfnamefont {C.}~\bibnamefont {Zhang}}, \ and\ \bibinfo {author} {\bibfnamefont {Y.~P.}\ \bibnamefont {Chen}},\ }\href {\doibase 10.1038/s41467-018-08119-4} {\bibfield  {journal} {\bibinfo  {journal} {Nat. Commun.}\ }\textbf {\bibinfo {volume} {10}},\ \bibinfo {pages} {375} (\bibinfo {year} {2019})}\BibitemShut {NoStop}%
\bibitem [{\citenamefont {Lin}\ \emph {et~al.}(2019)\citenamefont {Lin}, \citenamefont {Huang}, \citenamefont {Zhang}, \citenamefont {Zhu},\ and\ \citenamefont {Wang}}]{Lin2019Apr-soc-appli}%
  \BibitemOpen
  \bibfield  {author} {\bibinfo {author} {\bibfnamefont {Z.}~\bibnamefont {Lin}}, \bibinfo {author} {\bibfnamefont {X.-J.}\ \bibnamefont {Huang}}, \bibinfo {author} {\bibfnamefont {D.-W.}\ \bibnamefont {Zhang}}, \bibinfo {author} {\bibfnamefont {S.-L.}\ \bibnamefont {Zhu}}, \ and\ \bibinfo {author} {\bibfnamefont {Z.~D.}\ \bibnamefont {Wang}},\ }\href {\doibase 10.1103/PhysRevA.99.043419} {\bibfield  {journal} {\bibinfo  {journal} {Phys. Rev. A}\ }\textbf {\bibinfo {volume} {99}},\ \bibinfo {pages} {043419} (\bibinfo {year} {2019})}\BibitemShut {NoStop}%
\bibitem [{\citenamefont {Zheng}\ \emph {et~al.}(2019)\citenamefont {Zheng}, \citenamefont {Lin}, \citenamefont {Zhang}, \citenamefont {Zhu},\ and\ \citenamefont {Wang}}]{Zheng2019Nov-soc-appli}%
  \BibitemOpen
  \bibfield  {author} {\bibinfo {author} {\bibfnamefont {Z.}~\bibnamefont {Zheng}}, \bibinfo {author} {\bibfnamefont {Z.}~\bibnamefont {Lin}}, \bibinfo {author} {\bibfnamefont {D.-W.}\ \bibnamefont {Zhang}}, \bibinfo {author} {\bibfnamefont {S.-L.}\ \bibnamefont {Zhu}}, \ and\ \bibinfo {author} {\bibfnamefont {Z.~D.}\ \bibnamefont {Wang}},\ }\href {\doibase 10.1103/PhysRevResearch.1.033102} {\bibfield  {journal} {\bibinfo  {journal} {Phys. Rev. Res.}\ }\textbf {\bibinfo {volume} {1}},\ \bibinfo {pages} {033102} (\bibinfo {year} {2019})}\BibitemShut {NoStop}%
\bibitem [{\citenamefont {Chen}\ \emph {et~al.}(2020)\citenamefont {Chen}, \citenamefont {Zhang}, \citenamefont {Zhang},\ and\ \citenamefont {Zhu}}]{Chen2020Jan-soc-appli}%
  \BibitemOpen
  \bibfield  {author} {\bibinfo {author} {\bibfnamefont {Y.-L.}\ \bibnamefont {Chen}}, \bibinfo {author} {\bibfnamefont {G.-Q.}\ \bibnamefont {Zhang}}, \bibinfo {author} {\bibfnamefont {D.-W.}\ \bibnamefont {Zhang}}, \ and\ \bibinfo {author} {\bibfnamefont {S.-L.}\ \bibnamefont {Zhu}},\ }\href {\doibase 10.1103/PhysRevA.101.013627} {\bibfield  {journal} {\bibinfo  {journal} {Phys. Rev. A}\ }\textbf {\bibinfo {volume} {101}},\ \bibinfo {pages} {013627} (\bibinfo {year} {2020})}\BibitemShut {NoStop}%
\bibitem [{\citenamefont {Shen}\ \emph {et~al.}(2020)\citenamefont {Shen}, \citenamefont {Zhang}, \citenamefont {Yan}, \citenamefont {Li},\ and\ \citenamefont {Zhu}}]{Shen2020Jan-soc-appli}%
  \BibitemOpen
  \bibfield  {author} {\bibinfo {author} {\bibfnamefont {X.}~\bibnamefont {Shen}}, \bibinfo {author} {\bibfnamefont {D.-W.}\ \bibnamefont {Zhang}}, \bibinfo {author} {\bibfnamefont {H.}~\bibnamefont {Yan}}, \bibinfo {author} {\bibfnamefont {Z.}~\bibnamefont {Li}}, \ and\ \bibinfo {author} {\bibfnamefont {S.-L.}\ \bibnamefont {Zhu}},\ }\href {\doibase 10.1103/PhysRevResearch.2.013037} {\bibfield  {journal} {\bibinfo  {journal} {Phys. Rev. Res.}\ }\textbf {\bibinfo {volume} {2}},\ \bibinfo {pages} {013037} (\bibinfo {year} {2020})}\BibitemShut {NoStop}%
\bibitem [{\citenamefont {Zhu}\ \emph {et~al.}(2020)\citenamefont {Zhu}, \citenamefont {Ke}, \citenamefont {Zhong},\ and\ \citenamefont {Lee}}]{Zhu2020Apr-soc-appli}%
  \BibitemOpen
  \bibfield  {author} {\bibinfo {author} {\bibfnamefont {B.}~\bibnamefont {Zhu}}, \bibinfo {author} {\bibfnamefont {Y.}~\bibnamefont {Ke}}, \bibinfo {author} {\bibfnamefont {H.}~\bibnamefont {Zhong}}, \ and\ \bibinfo {author} {\bibfnamefont {C.}~\bibnamefont {Lee}},\ }\href {\doibase 10.1103/PhysRevResearch.2.023043} {\bibfield  {journal} {\bibinfo  {journal} {Phys. Rev. Res.}\ }\textbf {\bibinfo {volume} {2}},\ \bibinfo {pages} {023043} (\bibinfo {year} {2020})}\BibitemShut {NoStop}%
\bibitem [{\citenamefont {S{\ifmmode\acute{a}\else\'{a}\fi}nchez-Baena}\ \emph {et~al.}(2020)\citenamefont {S{\ifmmode\acute{a}\else\'{a}\fi}nchez-Baena}, \citenamefont {Boronat},\ and\ \citenamefont {Mazzanti}}]{Sanchez-Baena2020Apr-soc-appli}%
  \BibitemOpen
  \bibfield  {author} {\bibinfo {author} {\bibfnamefont {J.}~\bibnamefont {S{\ifmmode\acute{a}\else\'{a}\fi}nchez-Baena}}, \bibinfo {author} {\bibfnamefont {J.}~\bibnamefont {Boronat}}, \ and\ \bibinfo {author} {\bibfnamefont {F.}~\bibnamefont {Mazzanti}},\ }\href {\doibase 10.1103/PhysRevA.101.043602} {\bibfield  {journal} {\bibinfo  {journal} {Phys. Rev. A}\ }\textbf {\bibinfo {volume} {101}},\ \bibinfo {pages} {043602} (\bibinfo {year} {2020})}\BibitemShut {NoStop}%
\bibitem [{\citenamefont {Chen}\ \emph {et~al.}(2022{\natexlab{a}})\citenamefont {Chen}, \citenamefont {Liu},\ and\ \citenamefont {Hu}}]{Chen2022Aug-soc-appli}%
  \BibitemOpen
  \bibfield  {author} {\bibinfo {author} {\bibfnamefont {X.-L.}\ \bibnamefont {Chen}}, \bibinfo {author} {\bibfnamefont {X.-J.}\ \bibnamefont {Liu}}, \ and\ \bibinfo {author} {\bibfnamefont {H.}~\bibnamefont {Hu}},\ }\href {\doibase 10.1103/PhysRevA.106.023302} {\bibfield  {journal} {\bibinfo  {journal} {Phys. Rev. A}\ }\textbf {\bibinfo {volume} {106}},\ \bibinfo {pages} {023302} (\bibinfo {year} {2022}{\natexlab{a}})}\BibitemShut {NoStop}%
\bibitem [{\citenamefont {Zhang}\ and\ \citenamefont {Liu}(2022)}]{Zhang2022Oct-soc-appli}%
  \BibitemOpen
  \bibfield  {author} {\bibinfo {author} {\bibfnamefont {L.}~\bibnamefont {Zhang}}\ and\ \bibinfo {author} {\bibfnamefont {X.-J.}\ \bibnamefont {Liu}},\ }\href {\doibase 10.1103/PRXQuantum.3.040312} {\bibfield  {journal} {\bibinfo  {journal} {PRX Quantum}\ }\textbf {\bibinfo {volume} {3}},\ \bibinfo {pages} {040312} (\bibinfo {year} {2022})}\BibitemShut {NoStop}%
\bibitem [{\citenamefont {Zhu}\ \emph {et~al.}(2022)\citenamefont {Zhu}, \citenamefont {Zheng}, \citenamefont {Palumbo},\ and\ \citenamefont {Wang}}]{Zhu2022Nov-soc-appli}%
  \BibitemOpen
  \bibfield  {author} {\bibinfo {author} {\bibfnamefont {Y.-Q.}\ \bibnamefont {Zhu}}, \bibinfo {author} {\bibfnamefont {Z.}~\bibnamefont {Zheng}}, \bibinfo {author} {\bibfnamefont {G.}~\bibnamefont {Palumbo}}, \ and\ \bibinfo {author} {\bibfnamefont {Z.~D.}\ \bibnamefont {Wang}},\ }\href {\doibase 10.1103/PhysRevLett.129.196602} {\bibfield  {journal} {\bibinfo  {journal} {Phys. Rev. Lett.}\ }\textbf {\bibinfo {volume} {129}},\ \bibinfo {pages} {196602} (\bibinfo {year} {2022})}\BibitemShut {NoStop}%
\bibitem [{\citenamefont {Li}\ and\ \citenamefont {Yi}(2023)}]{Li2023Jan-soc-appli}%
  \BibitemOpen
  \bibfield  {author} {\bibinfo {author} {\bibfnamefont {H.}~\bibnamefont {Li}}\ and\ \bibinfo {author} {\bibfnamefont {W.}~\bibnamefont {Yi}},\ }\href {\doibase 10.1103/PhysRevA.107.013306} {\bibfield  {journal} {\bibinfo  {journal} {Phys. Rev. A}\ }\textbf {\bibinfo {volume} {107}},\ \bibinfo {pages} {013306} (\bibinfo {year} {2023})}\BibitemShut {NoStop}%
\bibitem [{\citenamefont {Xiong}\ \emph {et~al.}(2023)\citenamefont {Xiong}, \citenamefont {Gong}, \citenamefont {Fang},\ and\ \citenamefont {Sun}}]{Xiong2023Dec-soc-appli}%
  \BibitemOpen
  \bibfield  {author} {\bibinfo {author} {\bibfnamefont {L.}~\bibnamefont {Xiong}}, \bibinfo {author} {\bibfnamefont {M.}~\bibnamefont {Gong}}, \bibinfo {author} {\bibfnamefont {Z.-X.}\ \bibnamefont {Fang}}, \ and\ \bibinfo {author} {\bibfnamefont {R.}~\bibnamefont {Sun}},\ }\href {\doibase 10.1088/0256-307X/40/12/127402} {\bibfield  {journal} {\bibinfo  {journal} {Chin. Phys. Lett.}\ }\textbf {\bibinfo {volume} {40}},\ \bibinfo {pages} {127402} (\bibinfo {year} {2023})}\BibitemShut {NoStop}%
\bibitem [{\citenamefont {Wang}\ \emph {et~al.}(2024)\citenamefont {Wang}, \citenamefont {Zhu}, \citenamefont {Zhu},\ and\ \citenamefont {Zheng}}]{Wang2024Aug-soc-appli}%
  \BibitemOpen
  \bibfield  {author} {\bibinfo {author} {\bibfnamefont {Q.-D.}\ \bibnamefont {Wang}}, \bibinfo {author} {\bibfnamefont {Y.-Q.}\ \bibnamefont {Zhu}}, \bibinfo {author} {\bibfnamefont {S.-L.}\ \bibnamefont {Zhu}}, \ and\ \bibinfo {author} {\bibfnamefont {Z.}~\bibnamefont {Zheng}},\ }\href {\doibase 10.1103/PhysRevA.110.023321} {\bibfield  {journal} {\bibinfo  {journal} {Phys. Rev. A}\ }\textbf {\bibinfo {volume} {110}},\ \bibinfo {pages} {023321} (\bibinfo {year} {2024})}\BibitemShut {NoStop}%
\bibitem [{\citenamefont {Campbell}\ \emph {et~al.}(2016)\citenamefont {Campbell}, \citenamefont {Price}, \citenamefont {Putra}, \citenamefont {Vald{\ifmmode\acute{e}\else\'{e}\fi}s-Curiel}, \citenamefont {Trypogeorgos},\ and\ \citenamefont {Spielman}}]{Campbell2016Mar-soc-spin-1}%
  \BibitemOpen
  \bibfield  {author} {\bibinfo {author} {\bibfnamefont {D.~L.}\ \bibnamefont {Campbell}}, \bibinfo {author} {\bibfnamefont {R.~M.}\ \bibnamefont {Price}}, \bibinfo {author} {\bibfnamefont {A.}~\bibnamefont {Putra}}, \bibinfo {author} {\bibfnamefont {A.}~\bibnamefont {Vald{\ifmmode\acute{e}\else\'{e}\fi}s-Curiel}}, \bibinfo {author} {\bibfnamefont {D.}~\bibnamefont {Trypogeorgos}}, \ and\ \bibinfo {author} {\bibfnamefont {I.~B.}\ \bibnamefont {Spielman}},\ }\href {\doibase 10.1038/ncomms10897} {\bibfield  {journal} {\bibinfo  {journal} {Nat. Commun.}\ }\textbf {\bibinfo {volume} {7}},\ \bibinfo {pages} {10897} (\bibinfo {year} {2016})}\BibitemShut {NoStop}%
\bibitem [{\citenamefont {Mai}\ \emph {et~al.}(2018)\citenamefont {Mai}, \citenamefont {Zhu}, \citenamefont {Li}, \citenamefont {Zhang},\ and\ \citenamefont {Zhu}}]{Mai2018Nov-soc-spin-1}%
  \BibitemOpen
  \bibfield  {author} {\bibinfo {author} {\bibfnamefont {X.-Y.}\ \bibnamefont {Mai}}, \bibinfo {author} {\bibfnamefont {Y.-Q.}\ \bibnamefont {Zhu}}, \bibinfo {author} {\bibfnamefont {Z.}~\bibnamefont {Li}}, \bibinfo {author} {\bibfnamefont {D.-W.}\ \bibnamefont {Zhang}}, \ and\ \bibinfo {author} {\bibfnamefont {S.-L.}\ \bibnamefont {Zhu}},\ }\href {\doibase 10.1103/PhysRevA.98.053619} {\bibfield  {journal} {\bibinfo  {journal} {Phys. Rev. A}\ }\textbf {\bibinfo {volume} {98}},\ \bibinfo {pages} {053619} (\bibinfo {year} {2018})}\BibitemShut {NoStop}%
\bibitem [{\citenamefont {Hou}\ \emph {et~al.}(2020)\citenamefont {Hou}, \citenamefont {Hu},\ and\ \citenamefont {Zhang}}]{Hou2020May-soc-spin-1}%
  \BibitemOpen
  \bibfield  {author} {\bibinfo {author} {\bibfnamefont {J.}~\bibnamefont {Hou}}, \bibinfo {author} {\bibfnamefont {H.}~\bibnamefont {Hu}}, \ and\ \bibinfo {author} {\bibfnamefont {C.}~\bibnamefont {Zhang}},\ }\href {\doibase 10.1103/PhysRevA.101.053613} {\bibfield  {journal} {\bibinfo  {journal} {Phys. Rev. A}\ }\textbf {\bibinfo {volume} {101}},\ \bibinfo {pages} {053613} (\bibinfo {year} {2020})}\BibitemShut {NoStop}%
\bibitem [{\citenamefont {Adhikari}(2021)}]{Adhikari2021Jan-soc-spin-1}%
  \BibitemOpen
  \bibfield  {author} {\bibinfo {author} {\bibfnamefont {S.~K.}\ \bibnamefont {Adhikari}},\ }\href {\doibase 10.1103/PhysRevA.103.L011301} {\bibfield  {journal} {\bibinfo  {journal} {Phys. Rev. A}\ }\textbf {\bibinfo {volume} {103}},\ \bibinfo {pages} {L011301} (\bibinfo {year} {2021})}\BibitemShut {NoStop}%
\bibitem [{\citenamefont {Cabedo}\ \emph {et~al.}(2021)\citenamefont {Cabedo}, \citenamefont {Claramunt},\ and\ \citenamefont {Celi}}]{Cabedo2021Sep-soc-spin-1}%
  \BibitemOpen
  \bibfield  {author} {\bibinfo {author} {\bibfnamefont {J.}~\bibnamefont {Cabedo}}, \bibinfo {author} {\bibfnamefont {J.}~\bibnamefont {Claramunt}}, \ and\ \bibinfo {author} {\bibfnamefont {A.}~\bibnamefont {Celi}},\ }\href {\doibase 10.1103/PhysRevA.104.L031305} {\bibfield  {journal} {\bibinfo  {journal} {Phys. Rev. A}\ }\textbf {\bibinfo {volume} {104}},\ \bibinfo {pages} {L031305} (\bibinfo {year} {2021})}\BibitemShut {NoStop}%
\bibitem [{\citenamefont {Chen}\ \emph {et~al.}(2022{\natexlab{b}})\citenamefont {Chen}, \citenamefont {Lyu}, \citenamefont {Xu},\ and\ \citenamefont {Zhang}}]{Chen2022Aug-soc-spin-1}%
  \BibitemOpen
  \bibfield  {author} {\bibinfo {author} {\bibfnamefont {Y.}~\bibnamefont {Chen}}, \bibinfo {author} {\bibfnamefont {H.}~\bibnamefont {Lyu}}, \bibinfo {author} {\bibfnamefont {Y.}~\bibnamefont {Xu}}, \ and\ \bibinfo {author} {\bibfnamefont {Y.}~\bibnamefont {Zhang}},\ }\href {\doibase 10.1088/1367-2630/ac7fb1} {\bibfield  {journal} {\bibinfo  {journal} {New J. Phys.}\ }\textbf {\bibinfo {volume} {24}},\ \bibinfo {pages} {073041} (\bibinfo {year} {2022}{\natexlab{b}})}\BibitemShut {NoStop}%
\bibitem [{\citenamefont {Su}\ \emph {et~al.}(2023)\citenamefont {Su}, \citenamefont {Hou},\ and\ \citenamefont {Zhang}}]{Su2023Feb-soc-spin-1}%
  \BibitemOpen
  \bibfield  {author} {\bibinfo {author} {\bibfnamefont {Y.}~\bibnamefont {Su}}, \bibinfo {author} {\bibfnamefont {J.}~\bibnamefont {Hou}}, \ and\ \bibinfo {author} {\bibfnamefont {C.}~\bibnamefont {Zhang}},\ }\href {\doibase 10.1103/PhysRevB.107.085410} {\bibfield  {journal} {\bibinfo  {journal} {Phys. Rev. B}\ }\textbf {\bibinfo {volume} {107}},\ \bibinfo {pages} {085410} (\bibinfo {year} {2023})}\BibitemShut {NoStop}%
\bibitem [{\citenamefont {Banger}\ \emph {et~al.}(2023)\citenamefont {Banger}, \citenamefont {Rajat}, \citenamefont {Roy},\ and\ \citenamefont {Gautam}}]{Banger2023Oct-soc-spin-1}%
  \BibitemOpen
  \bibfield  {author} {\bibinfo {author} {\bibfnamefont {P.}~\bibnamefont {Banger}}, \bibinfo {author} {\bibnamefont {Rajat}}, \bibinfo {author} {\bibfnamefont {A.}~\bibnamefont {Roy}}, \ and\ \bibinfo {author} {\bibfnamefont {S.}~\bibnamefont {Gautam}},\ }\href {\doibase 10.1103/PhysRevA.108.043310} {\bibfield  {journal} {\bibinfo  {journal} {Phys. Rev. A}\ }\textbf {\bibinfo {volume} {108}},\ \bibinfo {pages} {043310} (\bibinfo {year} {2023})}\BibitemShut {NoStop}%
\bibitem [{\citenamefont {Xu}\ \emph {et~al.}(2024)\citenamefont {Xu}, \citenamefont {Chen},\ and\ \citenamefont {Chen}}]{Xu2024Jun-soc-spin-1}%
  \BibitemOpen
  \bibfield  {author} {\bibinfo {author} {\bibfnamefont {Y.}~\bibnamefont {Xu}}, \bibinfo {author} {\bibfnamefont {Y.}~\bibnamefont {Chen}}, \ and\ \bibinfo {author} {\bibfnamefont {X.}~\bibnamefont {Chen}},\ }\href {\doibase 10.1103/PhysRevA.109.063310} {\bibfield  {journal} {\bibinfo  {journal} {Phys. Rev. A}\ }\textbf {\bibinfo {volume} {109}},\ \bibinfo {pages} {063310} (\bibinfo {year} {2024})}\BibitemShut {NoStop}%
\bibitem [{\citenamefont {Gangwar}\ \emph {et~al.}(2024)\citenamefont {Gangwar}, \citenamefont {Ravisankar}, \citenamefont {Fabrelli}, \citenamefont {Muruganandam},\ and\ \citenamefont {Mishra}}]{Gangwar2024Apr-soc-spin-1}%
  \BibitemOpen
  \bibfield  {author} {\bibinfo {author} {\bibfnamefont {S.~K.}\ \bibnamefont {Gangwar}}, \bibinfo {author} {\bibfnamefont {R.}~\bibnamefont {Ravisankar}}, \bibinfo {author} {\bibfnamefont {H.}~\bibnamefont {Fabrelli}}, \bibinfo {author} {\bibfnamefont {P.}~\bibnamefont {Muruganandam}}, \ and\ \bibinfo {author} {\bibfnamefont {P.~K.}\ \bibnamefont {Mishra}},\ }\href {\doibase 10.1103/PhysRevA.109.043306} {\bibfield  {journal} {\bibinfo  {journal} {Phys. Rev. A}\ }\textbf {\bibinfo {volume} {109}},\ \bibinfo {pages} {043306} (\bibinfo {year} {2024})}\BibitemShut {NoStop}%
\bibitem [{\citenamefont {Bychkov}\ and\ \citenamefont {Rashba}(1984)}]{Bychkov1984Nov-rashba-soc}%
  \BibitemOpen
  \bibfield  {author} {\bibinfo {author} {\bibfnamefont {Y.~A.}\ \bibnamefont {Bychkov}}\ and\ \bibinfo {author} {\bibfnamefont {E.~I.}\ \bibnamefont {Rashba}},\ }\href {\doibase 10.1088/0022-3719/17/33/015} {\bibfield  {journal} {\bibinfo  {journal} {J. Phys. C: Solid State Phys.}\ }\textbf {\bibinfo {volume} {17}},\ \bibinfo {pages} {6039} (\bibinfo {year} {1984})}\BibitemShut {NoStop}%
\bibitem [{\citenamefont {Dresselhaus}(1955)}]{Dresselhaus1955Oct-dresselhaus-soc}%
  \BibitemOpen
  \bibfield  {author} {\bibinfo {author} {\bibfnamefont {G.}~\bibnamefont {Dresselhaus}},\ }\href {\doibase 10.1103/PhysRev.100.580} {\bibfield  {journal} {\bibinfo  {journal} {Phys. Rev.}\ }\textbf {\bibinfo {volume} {100}},\ \bibinfo {pages} {580} (\bibinfo {year} {1955})}\BibitemShut {NoStop}%
\bibitem [{\citenamefont {Qi}\ \emph {et~al.}(2008)\citenamefont {Qi}, \citenamefont {Hughes},\ and\ \citenamefont {Zhang}}]{Qi2008Nov-topo-ins}%
  \BibitemOpen
  \bibfield  {author} {\bibinfo {author} {\bibfnamefont {X.-L.}\ \bibnamefont {Qi}}, \bibinfo {author} {\bibfnamefont {T.~L.}\ \bibnamefont {Hughes}}, \ and\ \bibinfo {author} {\bibfnamefont {S.-C.}\ \bibnamefont {Zhang}},\ }\href {\doibase 10.1103/PhysRevB.78.195424} {\bibfield  {journal} {\bibinfo  {journal} {Phys. Rev. B}\ }\textbf {\bibinfo {volume} {78}},\ \bibinfo {pages} {195424} (\bibinfo {year} {2008})}\BibitemShut {NoStop}%
\bibitem [{\citenamefont {Gl{\ifmmode\ddot{u}\else\"{u}\fi}ck}\ \emph {et~al.}(2002)\citenamefont {Gl{\ifmmode\ddot{u}\else\"{u}\fi}ck}, \citenamefont {R.~Kolovsky},\ and\ \citenamefont {Korsch}}]{Gluck2002Aug-wannier-stark-ladder}%
  \BibitemOpen
  \bibfield  {author} {\bibinfo {author} {\bibfnamefont {M.}~\bibnamefont {Gl{\ifmmode\ddot{u}\else\"{u}\fi}ck}}, \bibinfo {author} {\bibfnamefont {A.}~\bibnamefont {R.~Kolovsky}}, \ and\ \bibinfo {author} {\bibfnamefont {H.~J.}\ \bibnamefont {Korsch}},\ }\href {\doibase 10.1016/S0370-1573(02)00142-4} {\bibfield  {journal} {\bibinfo  {journal} {Phys. Rep.}\ }\textbf {\bibinfo {volume} {366}},\ \bibinfo {pages} {103} (\bibinfo {year} {2002})}\BibitemShut {NoStop}%
\bibitem [{\citenamefont {Walters}\ \emph {et~al.}(2013)\citenamefont {Walters}, \citenamefont {Cotugno}, \citenamefont {Johnson}, \citenamefont {Clark},\ and\ \citenamefont {Jaksch}}]{Walters2013Apr-optical-lattice-cal}%
  \BibitemOpen
  \bibfield  {author} {\bibinfo {author} {\bibfnamefont {R.}~\bibnamefont {Walters}}, \bibinfo {author} {\bibfnamefont {G.}~\bibnamefont {Cotugno}}, \bibinfo {author} {\bibfnamefont {T.~H.}\ \bibnamefont {Johnson}}, \bibinfo {author} {\bibfnamefont {S.~R.}\ \bibnamefont {Clark}}, \ and\ \bibinfo {author} {\bibfnamefont {D.}~\bibnamefont {Jaksch}},\ }\href {\doibase 10.1103/PhysRevA.87.043613} {\bibfield  {journal} {\bibinfo  {journal} {Phys. Rev. A}\ }\textbf {\bibinfo {volume} {87}},\ \bibinfo {pages} {043613} (\bibinfo {year} {2013})}\BibitemShut {NoStop}%
\bibitem [{\citenamefont {Cohen-Tannoudji}\ \emph {et~al.}(1998)\citenamefont {Cohen-Tannoudji}, \citenamefont {Dupont-Roc},\ and\ \citenamefont {Grynberg}}]{Cohen-Tannoudji1998book}%
  \BibitemOpen
  \bibfield  {author} {\bibinfo {author} {\bibfnamefont {C.}~\bibnamefont {Cohen-Tannoudji}}, \bibinfo {author} {\bibfnamefont {J.}~\bibnamefont {Dupont-Roc}}, \ and\ \bibinfo {author} {\bibfnamefont {G.}~\bibnamefont {Grynberg}},\ }\href {\doibase 10.1002/9783527617197} {\emph {\bibinfo {title} {{Atom{\ifmmode---\else\textemdash\fi}Photon Interactions}}}}\ (\bibinfo  {publisher} {Wiley Online Library},\ \bibinfo {year} {1998})\BibitemShut {NoStop}%
\end{thebibliography}%
\end{document}